\documentclass{article}

\PassOptionsToPackage{numbers}{natbib}
\usepackage[final]{nips_2016}

\usepackage[utf8]{inputenc} % allow utf-8 input
\usepackage[T1]{fontenc}    % use 8-bit T1 fonts
\usepackage{url}            % simple URL typesetting
\usepackage{booktabs}       % professional-quality tables
\usepackage{amsfonts}       % blackboard math symbols
\usepackage{nicefrac}       % compact symbols for 1/2, etc.
\usepackage{microtype}      % microtypography

\usepackage{amsmath}
\usepackage[ruled, vlined, linesnumbered]{algorithm2e}
\providecommand{\SetAlgoLined}{\SetLine}
\usepackage{graphicx}
\usepackage{mathrsfs}
\usepackage{enumerate}
\usepackage{multirow}
\usepackage{color}
\usepackage{enumitem}
\usepackage{rotating}
\usepackage{url}
\usepackage{adjustbox}

\graphicspath{{./figures/}}

\newcommand{\argmax}{\operatornamewithlimits{argmax}}
\newcommand{\argmin}{\operatornamewithlimits{argmin}}
\newcommand{\diag}{\operatorname{diag}}
\newcommand{\vectorize}{\operatorname{vec}}
\newcommand{\tr}{\operatorname{tr}}

\title{Sparse Estimation of Multivariate \\ Poisson Log-Normal Models from Count
Data}

\author{
  Hao Wu \\
  Department of Electrical \\ and Computer Engineering\\
  Virginia Tech \\
  Arlington, VA 22203, USA \\
  \texttt{wuhao723@vt.edu} \\
  \And
  Xinwei Deng \\
  Department of Statistics \\
  Virginia Tech \\
  Blacksburg, VA 24061, USA\\
  \texttt{xdeng@vt.edu} \\
  \And
  Naren Ramakrishnan \\
  Department of Computer Science \\
  Virginia Tech \\
  Arlington, VA 22203, USA \\
  \texttt{naren@vt.edu} \\
}

\begin{document}

\maketitle

\begin{abstract}
Modeling data with multivariate count responses is a challenging problem due to
the discrete nature of the responses. Existing methods for univariate count
responses cannot be easily extended to the multivariate case since the
dependency among multiple responses needs to be properly accommodated. In this
paper, we propose a multivariate Poisson log-normal regression model for
multivariate data with count responses. By simultaneously estimating the
regression coefficients and inverse covariance matrix over the latent variables
with an efficient Monte Carlo EM algorithm, the proposed regression model takes
advantages of association among multiple count responses to improve the model
prediction performance. Simulation studies and applications to real world data
are conducted to systematically evaluate the performance of the proposed
method in comparison with conventional methods.
\end{abstract}

\section{Introduction}
\label{sec:intro}

In this decade of data science, multivariate response observations are routine
in numerous disciplines. To model such datasets, multivariate regression and
multi-task learning models are common techniques to study and investigate the
relationships between $q\ge 2$ responses and $p$ predictors. The former class of
methods,
e.g.~\cite{Rothman.2010.09188,Lozano:2013:RSE:2487575.2487667,NIPS2014_5630}
and~\cite{conf:aistats:WangLX13}, estimates the $p \times q$ regression
coefficients as well as recover the correlation structures among response
variables using regularization. The latter class of methods focuses on learning
the shared features~\cite{NIPS2012_4729,
Gong:2014:EMF:2623330.2623641,Gong:2012:RMF:2339530.2339672,NIPS2010_4125} or
common underlying structure(s) among multiple
tasks~\cite{NIPS2006_3143,conf:icml:KumarD12,Chen:2011:ILG:2020408.2020423,
Yu:2007:RML:1273496.1273635,conf:nips:ArgyriouMPY07} using regression approaches
and enforcing regularization controls over the coefficient matrix. However, all
such multivariate regression or multi-task learning models discussed above deal
with continuous responses, none of them handle count data.

When responses are count variables, the Poisson model is a natural approach to
model them, e.g., in domains such as influenza case count
modeling~\cite{Wang:2015:DPA:2783258.2783291}, traffic accident
analysis~\cite{Ma2008:MVPLN_crash,ElBasyouny2009820} and consumer
services~\cite{Deanna:Wang2007369}. However, Poisson regression models proposed
in these works are either univariate or inferred via Bayesian approaches and no
sparsity or feature selection is typically enforced over the coefficients.  When
count responses are multivariate, it is challenging to quantify the association
among them due to the discrete nature of the data. One important approach is to
model each dimension of count variables as the sum of independent Poisson
variables with some common Poisson variables capturing
dependencies~\cite{DKarlis2003}. A drawback of this method is that it can only
model positive correlations.  Recent literature~\cite{NIPS2013_5153,Hadiji2015}
models multivariate count data with novel Poisson graphical models which can
handle both positive and negative dependencies. However, these works do not
consider multivariate count data in the context of regression. 

To consider a joint model for data with multivariate count responses, it is
important to properly exploit the hidden associations among the count responses.
One way to consider the joint model of multivariate count responses is via
penalty-based model selection from the perspective of parameter regularization.
The key idea is to allow the count responses to be independent of each other,
while the regression coefficients are required to obey a certain common sparse
structure. Hence the joint modeling is enabled because of the joint estimation
of regression coefficients through appropriate penalties. Such a modeling
strategy leads to an explicit loss function with tractable computational
characteristics. However, this method overlooks the essential correlation among
multiple count responses, which could result in poor prediction performance.
There are also several recent papers that develop models of multivariate count
data from the lens of conditional dependency. But these method typically are
restricted to approximated likelihood functions under the framework of
generalized liner models. 

In this work, we propose a novel multivariate Poisson log-normal model for data
with multiple count responses. The motivation to adopt the log-normal model is
to borrow strength from regression under the multivariate normal assumption,
which can simultaneously estimate regression coefficients and covariance
structure.  For the proposed model, the logarithm of the Poisson rate parameters
is modeled as multivariate normal with a sparse inverse covariance matrix, which
combines the strengths of sparse regression and graphical modeling to improve
prediction performance. Thus, this approach can fully exploit the conditional
dependency among multiple count responses. Estimating such model is non-trivial
since it is intractable to derive an explicit analytical solution. Thus, to
facilitate the estimation of model parameters, we develop an Monte Carlo EM
algorithm which allows to iteratively estimate the regression coefficients using
the Lasso penalty and the inverse covariance matrix by a graphical Lasso
approach. By applying the proposed model to synthetic data and a real world
influenza-like illness dataset, we demonstrate the effectiveness of the proposed
method when modeling multivariate data with count responses.

It is worth pointing out that the proposed method is not restricted to adopt the
Lasso penalty for regression parameters. It can be easily extended to other
penalties such as the adaptive Lasso, group Lasso or fused Lasso. While
covariance matrix estimation and inverse covariance matrix estimation have
attracted significant attention in the
literature~\cite{Friedman:2134265,Rothman.2010.09188,Lozano:2013:RSE:2487575.2487667},
here we use this idea in the context of a multivariate regression for count
data. Thus inverse covariance matrix estimation is conducted here to improve
prediction performance, not just as an unsupervised procedure. One may call such
a strategy {\it supervised covariance estimation}, which has not been widely
studied in the literature. One exception is the multivariate regression for
continuous responses~\cite{Rothman.2010.09188,WytockK:icml13}. Therefore, to the
best of our knowledge, our proposed method is a first to incorporate covariance
matrix estimation into a multivariate regression model of count responses.

\section{Multivariate Poisson Log-Normal model}
\label{sec:MVPLN}

In this section, we formally specify the Multivariate Poisson Log-Normal (MVPLN)
model, and propose a Monte Carlo Expectation-Maximization (MCEM)
algorithm for parameter estimation in detail. 

\subsection{The proposed model}
\label{sec:spec}
\begin{figure}[!t]
	\centering
	\includegraphics[width=2in]{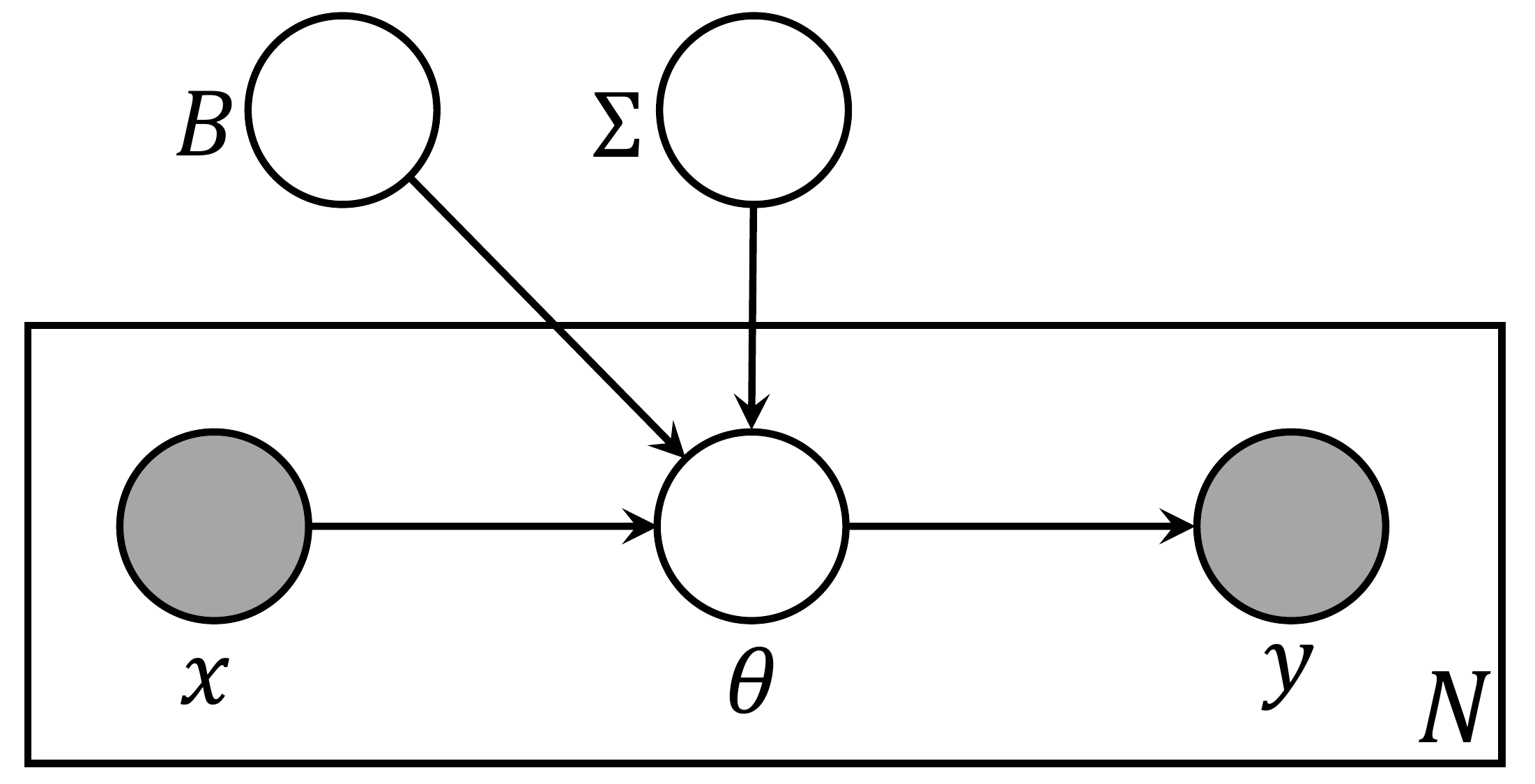}
	\caption{The plate notation of the proposed MVPLN model.}
	\label{fig:plate}
\end{figure}

Consider the multivariate random variable $\boldsymbol{\mathcal{Y}} =
\{\mathcal{Y}^{(1)}, $ $\mathcal{Y}^{(2)}, \ldots, \mathcal{Y}^{(q)} \}^T \in
\mathcal{Z}^{q}_{+}$, where the superscript $T$ denotes the transpose, and
$\mathcal{Z}_{+}$ represents the set of all positive integers. For count
data, it is reasonable to make the assumption that $\boldsymbol{\mathcal{Y}}$
follows the multivariate Poisson distribution. Without loss of generality, let's
assume that each dimension of $\boldsymbol{\mathcal{Y}}$, say
$\mathcal{Y}^{(i)}$, follows the univariate Poisson distribution with parameter
$\theta^{(i)}$, and is conditional independent of other dimensions given
$\theta^{(i)}$. That is:
\begin{align}
	\mathcal{Y}^{(i)} \sim
	\mathit{Poisson}\left(\theta^{(i)}\right),~\theta^{(i)} \in
	\mathcal{R}_{+},~\forall i = 1, 2,\ldots, q
	\label{eq:3.1.1}
\end{align}
Let $\boldsymbol{x} = {\{x^{(1)}, x^{(2)}, \ldots, x^{(p)}\}}^T \in
\mathcal{R}^{p}$ denote the predictor vector. In order to establish relationship
between $\boldsymbol{\mathcal{Y}}$ and $\boldsymbol{x}$, we consider the
following regression model:
\begin{align}
	\label{eq:3.1.2}
	\boldsymbol{\theta} = & \exp\left( \boldsymbol{B}^T \boldsymbol{x} +
	\boldsymbol{\varepsilon} \right)  \\
	\boldsymbol{\varepsilon} \sim & N(0, \boldsymbol{\Sigma}) \nonumber
\end{align}
where $\boldsymbol{B}$ is a $p \times q$ coefficient matrix, and
$\boldsymbol{\Sigma}$ is the $q \times q$ covariance matrix which captures the
covariance structure of variable $\boldsymbol{\theta} = {\{\theta^{(1)},
\theta^{(2)}, \ldots, \theta^{(q)}\}}^T$ given $\boldsymbol{x}$. Through the
variable $\boldsymbol{\theta}$, we model the covariance structure of the count
variable $\boldsymbol{\mathcal{Y}}$ indirectly. Fig.~\ref{fig:plate} shows the
plate notation of the proposed MVPLN model.

Given $n$ observations of the predictor $\boldsymbol{X} = [\boldsymbol{x}_1,
\boldsymbol{x}_2, $ ${\ldots, \boldsymbol{x}_n]}^T$ and corresponding responses
$\boldsymbol{Y} = [\boldsymbol{y}_1, \boldsymbol{y}_2, $ ${\ldots,
\boldsymbol{y}_n]}^T$, the log-likelihood of the MVPLN model is:
\begin{align}
	& \mathcal{L}(\boldsymbol{B}, \boldsymbol{\Sigma}) = \sum_{j=1}^{n} \log
	p(\boldsymbol{\mathcal{Y}} = \boldsymbol{y}_j \mid \boldsymbol{x}_j),
	\label{eq:3.2.5}
\end{align}
where
\begin{align}
	& p(\boldsymbol{\mathcal{Y}} = \boldsymbol{y} \mid \boldsymbol{x}) =
		\int_{\boldsymbol{\theta}} p(\boldsymbol{\mathcal{Y}} =
		\boldsymbol{y}, \boldsymbol{\theta} \mid \boldsymbol{x}) d
		\boldsymbol{\theta}  = \int_{\boldsymbol{\theta}}
		p(\boldsymbol{\mathcal{Y}} = \boldsymbol{y} \mid \boldsymbol{\theta})
		p(\boldsymbol{\theta} \mid \boldsymbol{x}) d \boldsymbol{\theta}
		\label{eq:3.2.4}
	%& p(\boldsymbol{\mathcal{Y}} = \boldsymbol{y}, \boldsymbol{\theta} \mid
	%\boldsymbol{x}) = p(\boldsymbol{\mathcal{Y}} = \boldsymbol{y} \mid
	%\boldsymbol{\theta}) p(\boldsymbol{\theta} \mid \boldsymbol{x})
	%\label{eq:3.2.3}
\end{align}
Here, $p(\boldsymbol{\mathcal{Y}} = \boldsymbol{y} \mid \boldsymbol{\theta})$
and $p(\boldsymbol{\theta} \mid \boldsymbol{x})$ follow multivariate Poisson
distribution and multivariate log-normal distribution as derived in Section A of
Supplementary Material, respectively. To jointly infer the sparse estimations of
coefficient matrix $\boldsymbol{B}$ and covariance matrix $\boldsymbol{\Sigma}$,
we adopt the regularized negative log-likelihood function with $l_1$ penalties
as our loss function. To be specific, the loss function could be written as:
\begin{align}
	\mathcal{L}_{p}(\boldsymbol{B}, \boldsymbol{\Sigma}) =
	-\mathcal{L}(\boldsymbol{B}, \boldsymbol{\Sigma}) + \lambda_1
	||\boldsymbol{B}||_{1} + \lambda_2 ||\boldsymbol{\Sigma}^{-1}||_{1},
	\label{eq:3.2.6}
\end{align}
where $||\cdot||_{1}$ denote the $l_{1}$ matrix norm, and $\lambda_1 > 0,
\lambda_2 > 0$ are two tuning parameters.

For convenience, we use the following notation to present the proposed MVPLN
model in the rest of the paper. Normal lower case letters, e.g.\ $x$ and $y$,
represent scalars. While, bold lower case letters, e.g.\ $\boldsymbol{x}$ and
$\boldsymbol{y}$, are used to represent column vectors, and bold upper case
letters in the calligraphic font, e.g.\ $\boldsymbol{\mathcal{X}}$ and
$\boldsymbol{\mathcal{Y}}$, denote random column vectors. Let letters with
superscript in parentheses, e.g.\ $x^{(i)}$, denote the $i^{th}$ component of
the corresponding vector $\boldsymbol{x}$. Matrices are represented by bold
upper case letters in normal font, e.g.\ $\boldsymbol{X}$ and $\boldsymbol{Y}$.
Letters in lower case with two subscripts, e.g.\ $x_{i,j}$, denote the $(i,j)$
entry of the corresponding matrix $\boldsymbol{X}$.

\subsection{Monte Carlo EM algorithm for parameter estimation}
\label{sec:em_algo}

In order to obtain the estimations of MVPLN model parameters $\boldsymbol{B}$
and $\boldsymbol{\Sigma}$, we could simply solve the following optimization
problem:
\begin{align}
	\hat{\boldsymbol{B}}, \hat{\boldsymbol{\Sigma}} = \argmin_{\boldsymbol{B},
	\boldsymbol{\Sigma}} \mathcal{L}_{p}(\boldsymbol{B}, \boldsymbol{\Sigma}).
	\label{eq:3.3.1}
\end{align}
However, it's difficult to directly minimize the objective function defined
above due to the complicated integral in Equation~\eqref{eq:3.2.4}. Thus, we
turn to an iterative approach for the solution. We treat $\boldsymbol{\theta}$
as latent random variables, and apply the EM algorithm to obtain the maximum
likelihood parameter estimation (MLE). However, we cannot derive the analytical
form of the expected log-likelihood of the model due to the integral in
Equation~\eqref{eq:3.2.4}. Here, we adopt a Monte Carlo variant of the EM
algorithm for an approximate solution.

\subsubsection{Monte Carlo (MC) E-step}
\label{sec:mc_estep}
%In the Monte Carlo (MC) E-step, instead of trying to derive the close form of
In the MC E-step of iteration $t + 1$, instead of trying to derive the close
form of the conditional probability distribution of $\boldsymbol{\theta}_j$, we
draw $m$ random samples of $\boldsymbol{\theta}_j$, say $\boldsymbol{\Theta}_j =
{\left[\boldsymbol{\theta}_j^{(1)}, \boldsymbol{\theta}_j^{(2)}, \ldots,
\boldsymbol{\theta}_j^{(m)}\right]}^T$, from $p(\boldsymbol{\theta}_j \mid
\boldsymbol{\mathcal{Y}} = \boldsymbol{y}_j, \boldsymbol{x}_j;
\boldsymbol{B}^{(t)}, \boldsymbol{\Sigma}^{(t)})$, and approximate the expected
log-likelihood function with:
\begin{align}
	\tilde{Q}(\boldsymbol{B}, & \boldsymbol{\Sigma} \mid \boldsymbol{B}^{(t)},
	\boldsymbol{\Sigma}^{(t)}) = \sum_{j=1}^{n} \frac{1}{m} \sum_{\tau =
	1}^{m} \log p(\boldsymbol{\mathcal{Y}} = \boldsymbol{y}_j,
	\boldsymbol{\theta}_j^{(\tau)} \mid \boldsymbol{x}_j; \boldsymbol{B}^{(t)},
	\boldsymbol{\Sigma}^{(t)}).
	\label{eq:3.4.1.1}
\end{align}
Drawing random samples of $\boldsymbol{\theta}_j$ can be achieved with the
Metropolis Hasting algorithm. In order to reduce the burn-in period of the
Metropolis Hasting algorithm, we adopt the tailored normal
distribution~\cite{Chib199833} as our proposal distribution. Since $
p(\boldsymbol{\theta}_j \mid \boldsymbol{\mathcal{Y}} = \boldsymbol{y}_j,
\boldsymbol{x}_j; \boldsymbol{B}^{(t)}, \boldsymbol{\Sigma}^{(t)}) \propto
p(\boldsymbol{\mathcal{Y}} = \boldsymbol{y}_j, \boldsymbol{\theta}_j \mid
\boldsymbol{x}_j; \boldsymbol{B}^{(t)}, \boldsymbol{\Sigma}^{(t)})$, if we let
$f(\boldsymbol{\theta}_j) = p(\boldsymbol{\mathcal{Y}} = \boldsymbol{y}_j,
\boldsymbol{\theta}_j \mid \boldsymbol{x}_j; \boldsymbol{B}^{(t)},
\boldsymbol{\Sigma}^{(t)})$, the initial value $\boldsymbol{\theta}_j^{(0)}$ of
the location parameter for the tailored normal distribution should be the mode
of $f(\boldsymbol{\theta}_j)$, and the covariance matrix is $\tau
{(-\boldsymbol{H}(\boldsymbol{\theta}_j^{(0)}))}^{-1}$,
where $\boldsymbol{H}(\boldsymbol{\theta}_j^{(0)})$ denotes the Hessian matrix
of $\log f(\boldsymbol{\theta}_j)$ at $\boldsymbol{\theta}_j^{(0)}$, and
$\tau$ is a tuning parameter. Considering the performance issue, we adopt a
linear approximation approach with the first order Taylor expansion to solve
$\boldsymbol{\theta}_j^{(0)}$. In this case, the approximate analytical solution
of $\boldsymbol{\theta}_j^{(0)}$ is $\boldsymbol{\theta}_j^{(0)} =
\boldsymbol{e}^{\boldsymbol{\kappa}_j}$ where
\begin{align*}
	\boldsymbol{\kappa}_j = &
	{\left( \diag \left(\boldsymbol{e}^{\boldsymbol{\kappa}_j^{(0)}}\right) +
	{\boldsymbol{\Sigma}^{(t)}}^{-1}\right)}^{-1} \Big( \boldsymbol{y}_j -
	\boldsymbol{1} + {\boldsymbol{\Sigma}^{(t)}}^{-1} {\boldsymbol{B}^{(t)}}^T
	\boldsymbol{x}_j +
	\diag\left(\boldsymbol{e}^{\boldsymbol{\kappa}_j^{(0)}}\right)
	\boldsymbol{\kappa}_j^{(0)} -
	\boldsymbol{e}^{\boldsymbol{\kappa}_j^{(0)}}\Big).	
\end{align*}
Here, $\boldsymbol{\kappa}_j^{(0)} = \log \boldsymbol{y}_j$. In the case that
the covariance matrix $\tau
{(-\boldsymbol{H}(\boldsymbol{\theta}_j^{(0)}))}^{-1}$ is not positive
semidefinite, the nearest positive semidefinite matrix to $\tau
{(-\boldsymbol{H}(\boldsymbol{\theta}_j^{(0)}))}^{-1}$ is used to replace $\tau
{(-\boldsymbol{H}(\boldsymbol{\theta}_j^{(0)}))}^{-1}$~\cite{Higham:nearPD}. The
details for Metropolis Hasting algorithm and the derivation of the tailored
normal distribution as the proposal distribution are provided in Section B of
the Supplementary Material.

\subsubsection{M-step: maximize approximate penalized expected log-likelihood}
\label{sec:m_step}
If we let $\boldsymbol{\Omega} = \boldsymbol{\Sigma}^{-1}$ and
$\boldsymbol{\varphi}_{\tau, j} = (\log \boldsymbol{\theta}_j^{(\tau)} -
\boldsymbol{B}^T \boldsymbol{x}_j)$, with the Monte Carlo approximation of the
expected log-likelihood in the MC E-step, the optimization problem we need to
solve in the M-step of the MCEM algorithm can be reformulated as:
\begin{align}
	\boldsymbol{B}^{(t+1)}, \boldsymbol{\Sigma}^{(t+1)} = &
	\argmin_{\boldsymbol{B}, \boldsymbol{\Omega}} \bigg\{ \frac{1}{mn}
	\tr\left(\boldsymbol{\Phi}^T \boldsymbol{\Phi} \boldsymbol{\Omega}\right) -
	\log |\boldsymbol{\Omega}| + \lambda_1 ||\boldsymbol{B}||_{1} + \lambda_2
	||\boldsymbol{\Omega}||_{1} \bigg\}
	\label{eq:3.4.2.3}
\end{align}
where $\boldsymbol{\Phi} = {[\boldsymbol{\varphi}_{1,1},
\boldsymbol{\varphi}_{2,1}, \ldots, \boldsymbol{\varphi}_{m,1},
\boldsymbol{\varphi}_{1,2}, \boldsymbol{\varphi}_{2,2}, \ldots,
\boldsymbol{\varphi}_{m,2}, \ldots, \boldsymbol{\varphi}_{m,n}]}^T$. The
optimization problem defined in Equation~\eqref{eq:3.4.2.3} is not convex.
However, it is convex w.r.t.\ either $\boldsymbol{B}$ or $\boldsymbol{\Omega}$
with the other fixed~\cite{Rothman.2010.09188}. Thus, we present an iterative
algorithm that optimizes the objective function in Equation~\eqref{eq:3.4.2.3}
alternatively w.r.t.\ $\boldsymbol{B}$ and $\boldsymbol{\Omega}$.

With $\boldsymbol{B}$ fixed at $\boldsymbol{B}_0$, the optimization problem in
Equation~\eqref{eq:3.4.2.3} yields:
\begin{align}
	\boldsymbol{\Omega}(\boldsymbol{B}_0) = \argmin_{\boldsymbol{\Omega}}
	\bigg\{ & \frac{1}{mn} \tr\left(\boldsymbol{\Phi}^T \boldsymbol{\Phi}
	\boldsymbol{\Omega} \right) - \log |\boldsymbol{\Omega}| + \lambda_2
	||\boldsymbol{\Omega}||_{1} \bigg\},
	\label{eq:3.4.2.4}
\end{align}
which is the similar problem studied in~\cite{Friedman:2134265}. We solve this
problem with the graphical lasso approach.

When $\boldsymbol{\Omega}$ is fixed at $\boldsymbol{\Omega}_0$, we have the
following optimization problem:
\begin{align}
	\boldsymbol{B}(\boldsymbol{\Omega}_0) = \argmin_{\boldsymbol{B}}
	\bigg\{ \frac{1}{mn} \tr\left( \boldsymbol{\Phi}^T \boldsymbol{\Phi}
	\boldsymbol{\Omega}_0 \right) + \lambda_1 ||\boldsymbol{B}||_{1} \bigg\},
	\label{eq:3.4.2.5}
\end{align}
which is similar to the problem solved by Lasso, and we could adopt the cyclical
coordinate descent algorithm~\cite{friedman2007:pathwise} to obtain the
estimation of $\boldsymbol{B}$. However, considering the computational burden
already brought in by the MCMC approximation in the MC E-step, we solve the
optimization problem in Equation~\eqref{eq:3.4.2.5} approximately where the
$l_1$ matrix norm $||\boldsymbol{B}||_{1}$ is replaced by its quadratic
approximation $\tr({\boldsymbol{B}'}^T \boldsymbol{B}')$ where $\boldsymbol{B}'
= \boldsymbol{B} \circ ({1}/{\sqrt{|\hat{\boldsymbol{B}}|}})$.  Here, $\circ$
denotes the Hadamard (element-wise) product, $\hat{\boldsymbol{B}}$ denotes the
current estimation of $\boldsymbol{B}$, and $1/\sqrt{|\hat{\boldsymbol{B}}|}$
represents the matrix that each entry is the inverse of the square root of the
absolute value of the corresponding entry in $\hat{\boldsymbol{B}}$. With such
approximation, we would get the analytical solution to the optimization problem
in Equation~\eqref{eq:3.4.2.5} as
\begin{align}
	\vectorize(\boldsymbol{B}) = {\left[\boldsymbol{\Omega}_0 \otimes
	\boldsymbol{S} + \diag\left(\vectorize\left(\frac{\lambda_1 m
	n}{|\hat{\boldsymbol{B}}|}\right)\right)\right]}^{-1} \hspace{-0.2cm}
	\vectorize(\boldsymbol{H}).
	\label{eq:3.4.2.8}
\end{align}
Here, $\vectorize(\cdot)$ represents the vectorization operation over the
matrix, and the two auxiliary matrices $\boldsymbol{H}$ and $\boldsymbol{S}$
are:
$$
\boldsymbol{H} = \Big(\sum\limits_{j=1}^{n} \boldsymbol{X}_j^T (\log
\boldsymbol{\Theta}_j)\Big) \boldsymbol{\Omega}_0, \quad \boldsymbol{S} =
\sum\limits_{j = 1}^{n} \boldsymbol{X}_j^T \boldsymbol{X}_j,
$$
where $\boldsymbol{X}_j$ is a $m \times p$ matrix with each row being
$\boldsymbol{x}_j$ for all $j = 1, 2, \ldots, n$. The estimated coefficient
matrix $\boldsymbol{B}$ can be obtained by reorganizing the
$\vectorize(\boldsymbol{B})$ in Equation~\eqref{eq:3.4.2.8}. By solving the
optimization problem in Equation~\eqref{eq:3.4.2.4} and~\eqref{eq:3.4.2.5}
alternatively until convergence, we will obtain the MLE of the coefficient
matrix $\boldsymbol{B}$ and inverse covariance matrix $\boldsymbol{\Omega}$. The
detailed derivation of the algorithm for M-step is provided in Section C of the
Supplementary Material.

\subsection{Selection of tuning parameters}
\label{sec:ebic}
To determine the optimal values of the tuning parameters $\lambda_1$ and
$\lambda_2$, we adopt the extended Bayesian Information Criterion (EBIC)
approach proposed in~\cite{Jiahua:v:95:y:2008:i:3} and extended to Gaussian
Graphical Models in~\cite{NIPS2010_4087}. Assume
$\boldsymbol{B}_{\lambda_1, \lambda_2}$ and $\boldsymbol{\Omega}_{\lambda_1,
\lambda_2}$ denote the MLE of the model parameter $\boldsymbol{B}$ and
$\boldsymbol{\Omega}$ with regularization parameters $\lambda_1$ and $\lambda_2$.
The EBIC value for this model is given by the following equation:
\begin{align}
	\text{EBIC}_{\gamma} (\lambda_1, \lambda_2) = & -2
	\tilde{Q}(\boldsymbol{B}_{\lambda_1, \lambda_2},
	\boldsymbol{\Omega}_{\lambda_1, \lambda_2})
	+ [v(\boldsymbol{B}_{\lambda_1, \lambda_2}) +
	v(\boldsymbol{\Omega}_{\lambda_1, \lambda_2})] \log n + 2 \gamma
	v(\boldsymbol{B}_{\lambda_1, \lambda_2}) \log (pq) \nonumber \\
	& + 4 \gamma v(\boldsymbol{\Omega}_{\lambda_1, \lambda_2}) \log q,
	\label{eq:3.5.1}
\end{align}
where $\tilde{\boldsymbol{Q}}(\boldsymbol{B}_{\lambda_1, \lambda_2},
\boldsymbol{\Omega}_{\lambda_1, \lambda_2})$ is the approximate expected
log-likelihood in Equation~\eqref{eq:3.4.1.1}, $v(\boldsymbol{B}_{\lambda_1,
\lambda_2})$ and $v(\boldsymbol{\Omega}_{\lambda_1, \lambda_2})$ denote the
number of non-zero entries in $\boldsymbol{B}_{\lambda_1, \lambda_2}$ and $
\boldsymbol{\Omega}_{\lambda_1, \lambda_2}$, respectively, and $n$ is the number
of training observations. With EBIC, the optimal values for $\lambda_1$ and
$\lambda_2$ are selected by
\begin{align*}
	(\hat{\lambda}_1, \hat{\lambda}_2) = \argmin_{\lambda_1, \lambda_2}
	\text{EBIC}_{\gamma}(\lambda_1, \lambda_2).
\end{align*}

\section{Experiments and results}
\label{sec:experiment}
\subsection{Simulation study}

In our simulation study, we compare the proposed MVPLN model with the separate
univariate Lasso regularized Poisson regression model (GLMNET model) (e.g., as
implemented in the R glmnet package~\cite{r:glmnet}). The regularized univariate
Poisson regression is applied to each response dimension, and a Bayesian
Information Criterion (BIC) is used to select regularization parameters in order
to make a fair comparison. The simulation data are generated with the following
approach. Each data observation in the $n \times p$ predictor matrix
$\boldsymbol{X}$ is independently sampled from a multivariate normal
distribution $N(\boldsymbol{\mu}_{X}, \sigma_{X} \boldsymbol{I})$, where the
location parameter $\boldsymbol{\mu}_{X}$ is sampled from a uniform distribution
$\mathit{Unif}(\boldsymbol{\mu}_{\min}, \boldsymbol{\mu}_{\max})$. The
corresponding observations in the $n \times q$ response matrix $\boldsymbol{Y}$
are generated following the definition of the MVPLN model in
Equation~\eqref{eq:3.1.1} and~\eqref{eq:3.1.2}. In order to enforce sparsity, a
fixed number of zeros are randomly placed into each column of the coefficient
matrix $\boldsymbol{B}$. The other non-zero entries of $\boldsymbol{B}$ are
independently sampled from a univariate normal distribution $N(\mu_{B},
\sigma_{B})$. Regarding the inverse covariance matrix $\boldsymbol{\Omega} =
\boldsymbol{\Sigma}^{-1}$ for $\boldsymbol{\varepsilon}$, we consider four
scenarios: (1).\ Random $\boldsymbol{\Omega}$, where the inverse covariance
matrix is generated by $\boldsymbol{\Omega} = \boldsymbol{\Psi}^T
\boldsymbol{\Psi}$ to ensure the positive semidefinite property. Each entry in
$\boldsymbol{\Psi}$ is independently sampled from a uniform distribution
$\mathit{Unif}(-1,1)$; (2).\ Banded $\boldsymbol{\Omega}$, where the sparsity is
enforced by the modified Cholesky decomposition~\cite{levina2008sparse}:
$\boldsymbol{\Omega} = \boldsymbol{T}^{T} \boldsymbol{D}^{-1} \boldsymbol{T}$.
Here, $\boldsymbol{T}$ is a lower triangular matrix with $1's$ on the diagonal,
and $\boldsymbol{D}$ is a diagonal matrix. The non-zero off diagonal elements in
$\boldsymbol{T}$ and diagonal elements in $\boldsymbol{D}$ are independently
sampled from uniform distribution $\mathit{Unif}(-1,1)$ and $\mathit{Unif}(0,1)$
respectively; (3).\ sparse $\boldsymbol{\Omega}$, where the
$\boldsymbol{\Omega}$ matrix is generated by performing some random row and
column permutations over the banded $\boldsymbol{\Omega}$ matrix; (4).\ Diagonal
$\boldsymbol{\Omega}$, where the diagonal elements are sampled independently
from standard uniform distribution. In order to make sure that the elements in
the response matrix $\boldsymbol{Y}$ are within the reasonable range, we scale
the matrix $\boldsymbol{\Sigma}$ to make the largest element equal to $\psi$. By
tuning the synthetic data generation parameters
$\boldsymbol{\mu}_{\min},~\boldsymbol{\mu}_{\max},~\sigma_X,~\mu_B,~\sigma_B$,
and $\psi$, we could adjust the range and variations in the generated response
matrix $\boldsymbol{Y}$.

In our experiments, we fix the number of observations in the training data at
$n=50$, and the number of observations in the test data at $20$. We consider two
scenarios: (1) the dimension of predictors is less than the number of
observations in training data ($p < n$); (2) the dimension of predictors is
greater than or equal to the number of observations in training data ($p \geq
n$). We let $p = 30, q = 5$ for the case $p < n$, and $p = 70, q = 5$ for the
case $p \geq n$. For each parameter setting, the simulation is repeated for $60$
times, and the reported results are averaged across the $60$ replications to
alleviate the randomness.
\begin{table*}[t]
	\centering
	\caption{Estimation errors w.r.t.\ $\boldsymbol{B}$ and
	$\boldsymbol{\Omega}$. The standard errors are shown in the parentheses.}
	\label{tab:est_error}
	\begin{adjustbox}{max width=\textwidth}
	\begin{tabular}{c c c c c c c c c c}
		\toprule
		\multirow{3}{*}[-5pt]{$\boldsymbol{\Omega}$} & \multirow{3}{*}[-5pt]{$\psi$} &
		\multicolumn{4}{c}{$l(\boldsymbol{B}, \hat{\boldsymbol{B}})$} &
		\multicolumn{4}{c}{$l(\boldsymbol{\Omega},
		\boldsymbol{\hat{\boldsymbol{\Omega}}})$} \\ \cmidrule{3-10}	
		& & \multicolumn{2}{c}{$p < n$} & \multicolumn{2}{c}{$p > n$} &
		\multicolumn{2}{c}{$p < n$} & \multicolumn{2}{c}{$p > n$}\\
		\cmidrule{3-10}
		& & GLMNET & MVPLN & GLMNET & MVPLN & GLMNET & MVPLN & GLMNET & MVPLN \\ 
		\midrule
		\multirow{8}{*}{Random} 
		& $0.4$ & 2.25607 & \textbf{1.19936} & 1.61016 & \textbf{1.44383} & NA & 0.99550 & NA & 0.99595 \\
		&       & (0.04547) & \textbf{(0.01277)} & (0.01830) & \textbf{(0.01076)} & & (0.00131) & & (0.00100) \\
		& $1.0$ & 4.35649 & \textbf{1.70326} & 2.41644 & \textbf{1.74861} & NA & 0.99033 & NA & 0.99151 \\
		&       & (0.09258) & \textbf{(0.03620)} & (0.03039) & \textbf{(0.02796)} & & (0.00153) & & (0.00200) \\
		& $1.6$ & 5.37513 & \textbf{1.80392} & 2.87839 & \textbf{1.94325} & NA & 0.98928 & NA & 0.98561 \\
		&       & (0.12519) & \textbf{(0.03618)} & (0.04629) & \textbf{(0.02844)} & & (0.00211) & & (0.00452) \\
		& $2.2$ & 6.32172 & \textbf{1.99852} & 3.21878 & \textbf{2.12487} & NA & 0.99214 & NA & 0.98343 \\ 
		&       & (0.17932) & \textbf{(0.04246)} & (0.05822) & \textbf{(0.04339)} & & (0.00126) & & (0.00328) \\
		\midrule
		\multirow{8}{*}{Banded} 
		& $0.4$ & 2.12650 & \textbf{1.16671} & 1.49028 & \textbf{1.38619} & NA & 0.98029 & NA & 0.98500 \\
		&       & (0.05377) & \textbf{(0.01882)} & (0.01747) & \textbf{(0.01133)} & & (0.00204) & & (0.00148) \\
		& $1.0$ & 3.57945 & \textbf{1.59062} & 2.13400 & \textbf{1.59255} & NA & 0.95796 & NA & 0.94881 \\
		&       & (0.10031) & \textbf{(0.04760)} & (0.03313) & \textbf{(0.02344)} & & (0.00508) & & (0.00563) \\
		& $1.6$ & 4.41182 & \textbf{1.80692} & 2.59768 & \textbf{1.78361} & NA & 0.93159 & NA & 0.92380 \\
		&       & (0.13408) & \textbf{(0.06746)} & (0.05930) & \textbf{(0.02981)} & & (0.00811) & & (0.00874) \\
		& $2.2$ & 5.21359 & \textbf{2.04397} & 2.84779 & \textbf{2.01992} & NA & 0.93695 & NA & 0.90552 \\
		&       & (0.18171) & \textbf{(0.07308)} & (0.07824) & \textbf{(0.05492)} & & (0.00681) & & (0.00838) \\
		\midrule
		\multirow{8}{*}{Sparse} 
		& $0.4$ & 1.98327 & \textbf{1.11950} & 1.52410 & \textbf{1.40847} & NA & 0.98277 & NA & 0.98205 \\
		&       & (0.06026) & \textbf{(0.01556)} & (0.02270) & \textbf{(0.01107)} & & (0.00259) & & (0.00201) \\
		& $1.0$ & 3.43339 & \textbf{1.50384} & 2.13721 & \textbf{1.60572} & NA & 0.95978 & NA & 0.96085 \\
		&       & (0.11127) & \textbf{(0.04915)} & (0.03966) & \textbf{(0.02315)} & & (0.00597) & & (0.00425) \\
		& $1.6$ & 4.69189 & \textbf{1.88319} & 2.54446 & \textbf{1.76144} & NA & 0.92684 & NA & 0.92349 \\
		&       & (0.15989) & \textbf{(0.07134)} & (0.05723) & \textbf{(0.02705)} & & (0.00880) & & (0.00852) \\
		& $2.2$ & 5.09710 & \textbf{2.12963} & 2.74681 & \textbf{1.91288} & NA & 0.96626 & NA & 0.90581 \\ 
		&       & (0.21733) & \textbf{(0.07617)} & (0.08444) & \textbf{(0.04085)} & & (0.01344) & & (0.00957) \\
		\midrule
		\multirow{8}{*}{Diagonal}
		& $0.4$ & 1.86103 & \textbf{1.10292} & 1.43937 & \textbf{1.34607} & NA & 0.96841 & NA & 0.97068 \\
		&       & (0.05898) & \textbf{(0.01452)} & (0.01870) & \textbf{(0.01274)} & & (0.00324) & & (0.00413) \\
		& $1.0$ & 3.29868 & \textbf{1.53224} & 2.01567 & \textbf{1.56539} & NA & 0.88673 & NA & 0.89628 \\
		&       & (0.09724) & \textbf{(0.04655)} & (0.04295) & \textbf{(0.02745)} & & (0.01313) & & (0.01510) \\
		& $1.6$ & 4.33160 & \textbf{1.84269} & 2.39551 & \textbf{1.70712} & NA & 0.81895 & NA & 0.84071 \\
		&       & (0.13345) & \textbf{(0.06302)} & (0.05794) & \textbf{(0.04889)} & & (0.01851) & & (0.02020) \\
		& $2.2$ & 5.00582 & \textbf{1.95903} & 2.56122 & \textbf{1.76716} & NA & 0.88405 & NA & 0.81663 \\ 
		&       & (0.23160) & \textbf{(0.08481)} & (0.08119) & \textbf{(0.03930)} & & (0.02031) & & (0.02034) \\
		\bottomrule
	\end{tabular}
	\end{adjustbox}
\end{table*}

%\subsubsection{Capability of Sparse Estimation}
%\label{sec:heatmap}
%To measure the abilities of variable selection and recovering the structures of
%coefficient matrix $\boldsymbol{B}$ and inverse covariance matrix
%$\boldsymbol{\Omega}$, we plot the heatmap of the estimation
%$\hat{\boldsymbol{B}}$ and $\hat{\boldsymbol{\Omega}}$. The heatmap for
%$\hat{\boldsymbol{B}}$ is defined as follow (similar for
%$\hat{\boldsymbol{\Omega}}$). For each entry in $\hat{\boldsymbol{B}}$, if the
%estimated coefficient is not zero, we replace it with $1$, otherwise, it is
%zero. Then, it is averaged across $60$ replications in our experiment, and
%compared to the true coefficient matrix $B$ used when generating the simulated
%data. Figure~\ref{fig:Omega_p30},~\ref{fig:B_p30} and~\ref{fig:B_p70} show the
%heatmaps for $\boldsymbol{\Omega}$ and $\boldsymbol{B}$ for the cases when $p <
%n$ and $p > n$ respectively. Since the GLMNET model does not have the capability
%of inverse covariance matrix estimation, we only show the heatmap of
%$\boldsymbol{\Omega}$ for the proposed MVPLN model.\haocomment{we need some
%explanation here for the heatmap.}

\subsubsection{Estimation accuracy}
\label{sec:accuracy}
To measure model estimation accuracy w.r.t.\ $\boldsymbol{B}$ and
$\boldsymbol{\Omega}$, we report the estimation errors by computing the distance
between $\boldsymbol{B}$ and $\hat{\boldsymbol{B}}$ (or $\boldsymbol{\Omega} =
\boldsymbol{\Sigma}^{-1}$ and $\hat{\boldsymbol{\Omega}} =
\hat{\boldsymbol{\Sigma}}^{-1}$) using the normalized matrix Frobenius norm:
\begin{align*}
	l(\boldsymbol{B}, \hat{\boldsymbol{B}}) = \frac{||\boldsymbol{B} -
	\hat{\boldsymbol{B}}||_{F}}{||\boldsymbol{B}||_{F}}
\end{align*}
Here, $\boldsymbol{B}$ denotes the true value of coefficient matrix and
$\hat{\boldsymbol{B}}$ represents the estimation given by the MVPLN or GLMNET
model. Table~\ref{tab:est_error} shows the estimation errors of coefficient
matrix $\boldsymbol{B}$ and inverse covariance matrix $\boldsymbol{\Omega}$ in
various parameter settings. Since the GLMNET model cannot infer the inverse
covariance matrix, we omit the corresponding results here. We can see that the
proposed MVPLN model consistently outperforms the GLMNET model in all parameter
settings, especially when the variation in the simulated data is large ($\psi$
is large). Such promising results demonstrate that the proposed MVPLN model
leverages the dependency structures between the multi-dimensional count
responses to improve the estimation accuracy.

\begin{figure*}[t]
	\hspace{-0.1in}
	\begin{minipage}[t]{0.24\linewidth}
		\centering
		\includegraphics[width=1.6in]{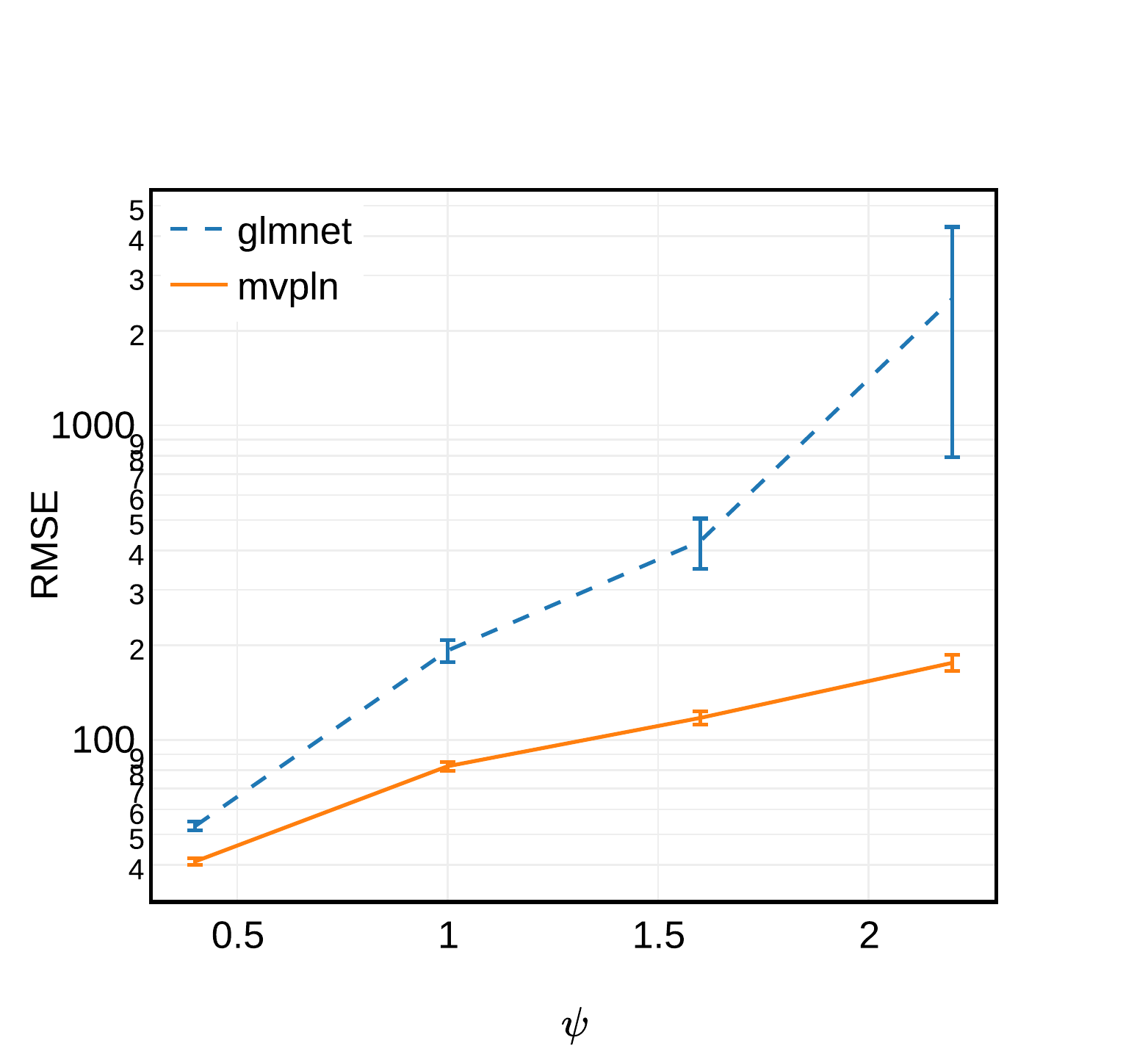}
	\end{minipage}
	\hspace{0.015in}
	\begin{minipage}[t]{0.24\linewidth}
		\centering
		\includegraphics[width=1.6in]{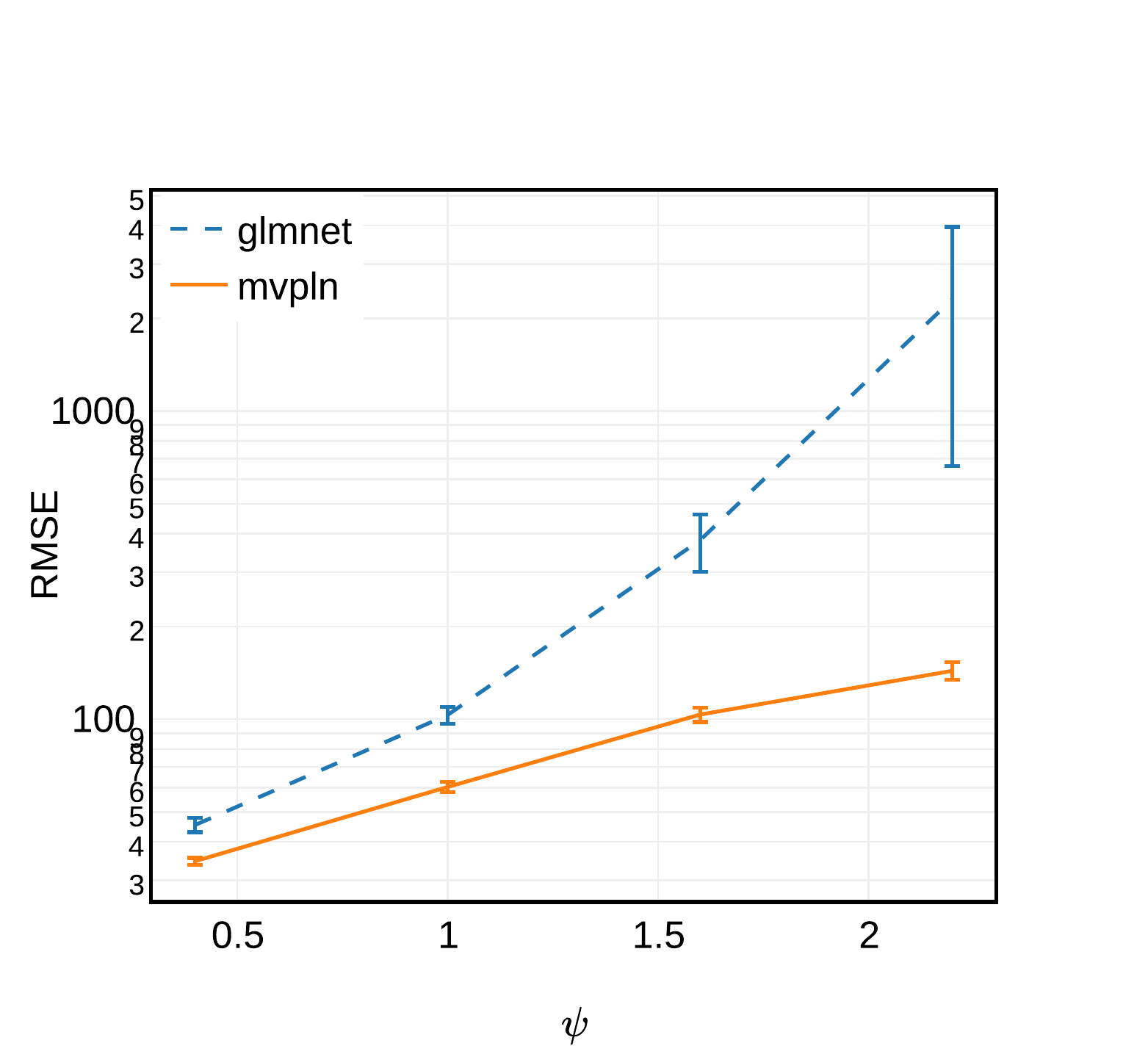}
	\end{minipage}
	\hspace{0.015in}
	\begin{minipage}[t]{0.24\linewidth}
		\centering
		\includegraphics[width=1.6in]{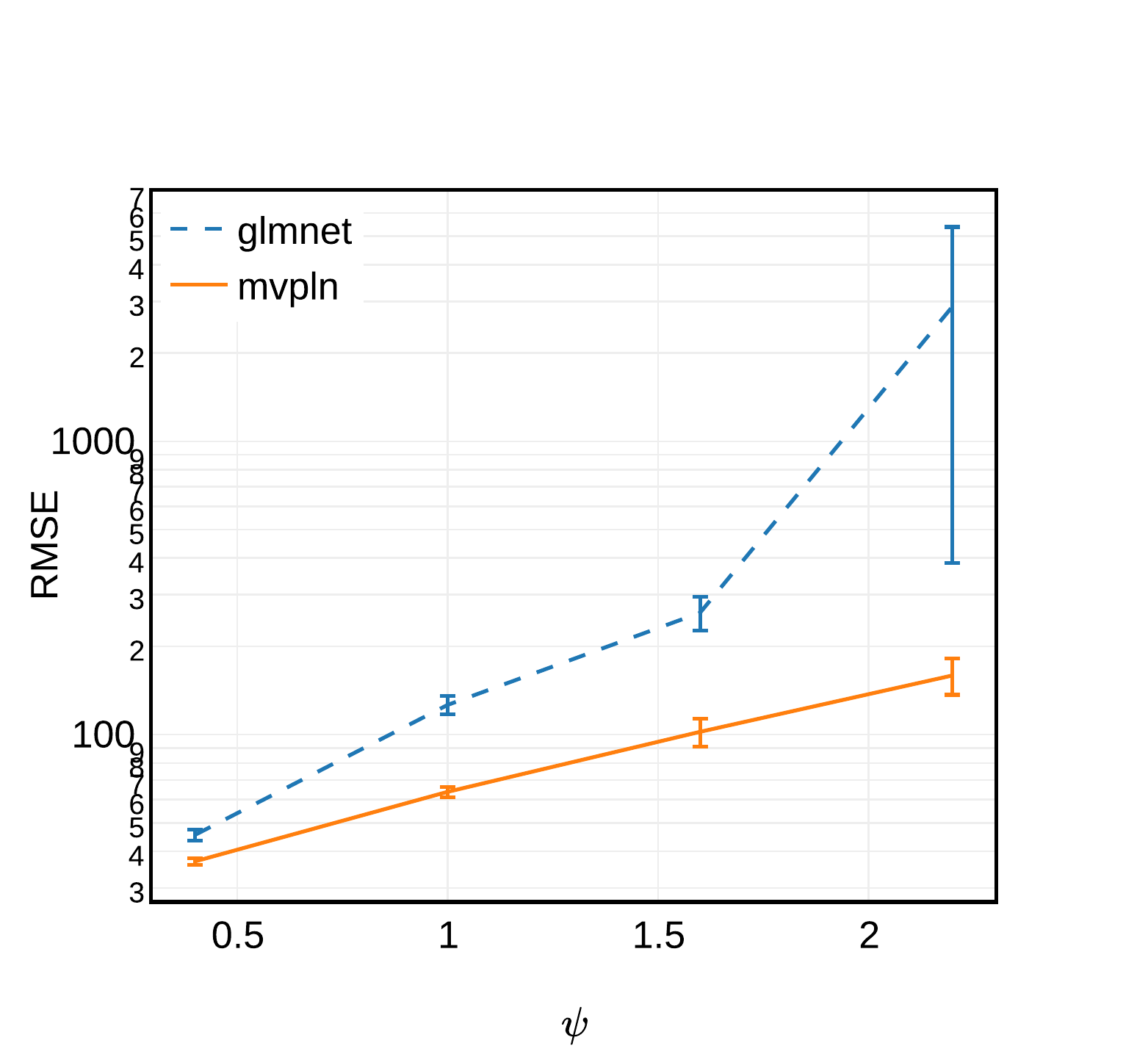}
	\end{minipage}
	\hspace{0.015in}
	\begin{minipage}[t]{0.24\linewidth}
		\centering
		\includegraphics[width=1.6in]{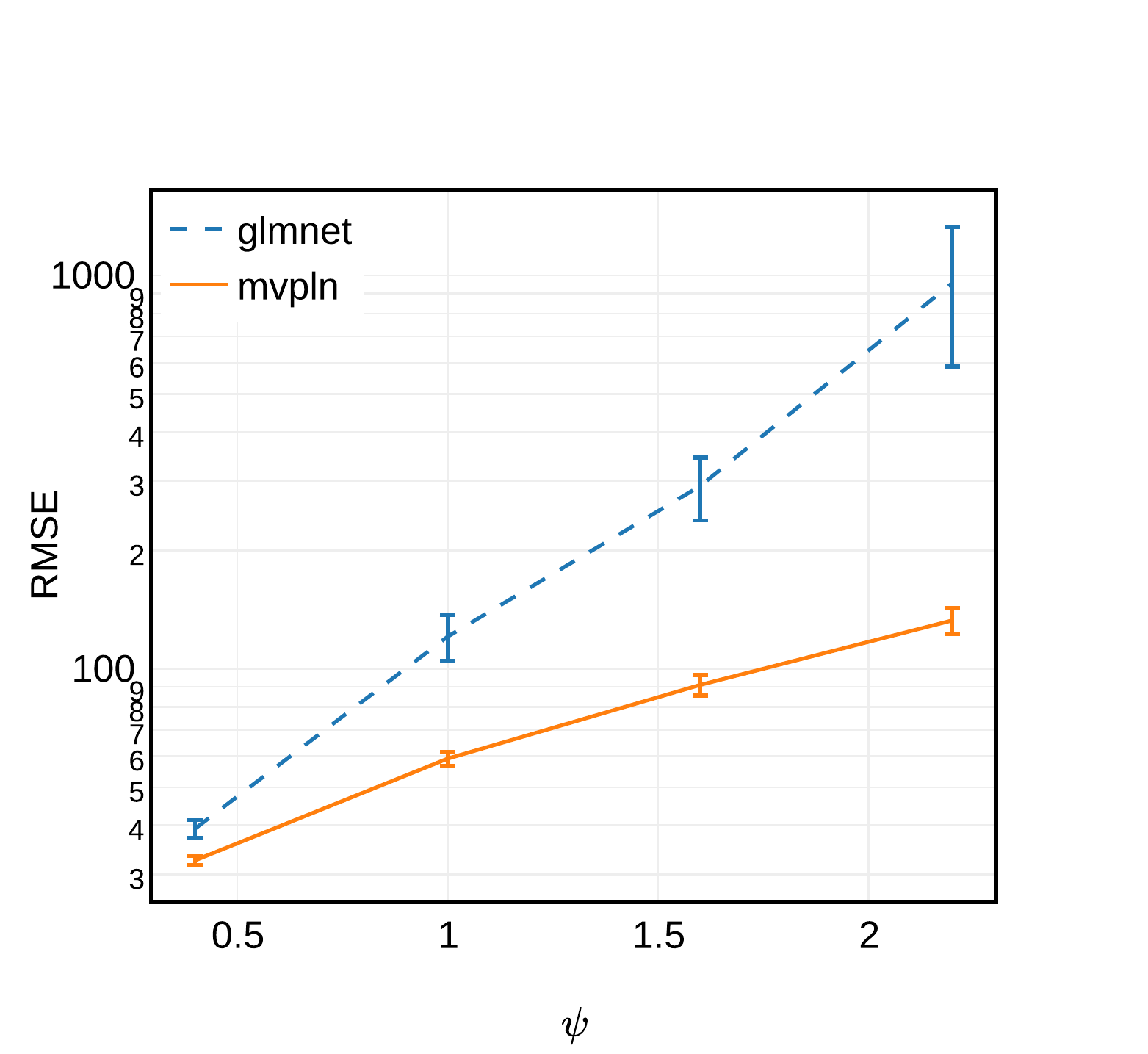}
	\end{minipage}
	\caption{Average rMSE across response dimensions over test data when
		$\boldsymbol{\Omega}$ is random, sparse, banded and diagonal
		(from left to right), and $p < n$. The vertical error bars indicate the
		standard deviation, and the Y-axis is in log scale.}\label{fig:4.1.1}
		%\vspace{-0.15in}
\end{figure*}

\begin{figure*}[t]
	%\hspace{-0.1in}
	\begin{minipage}[t]{0.24\linewidth}
		\centering
		\includegraphics[width=1.6in]{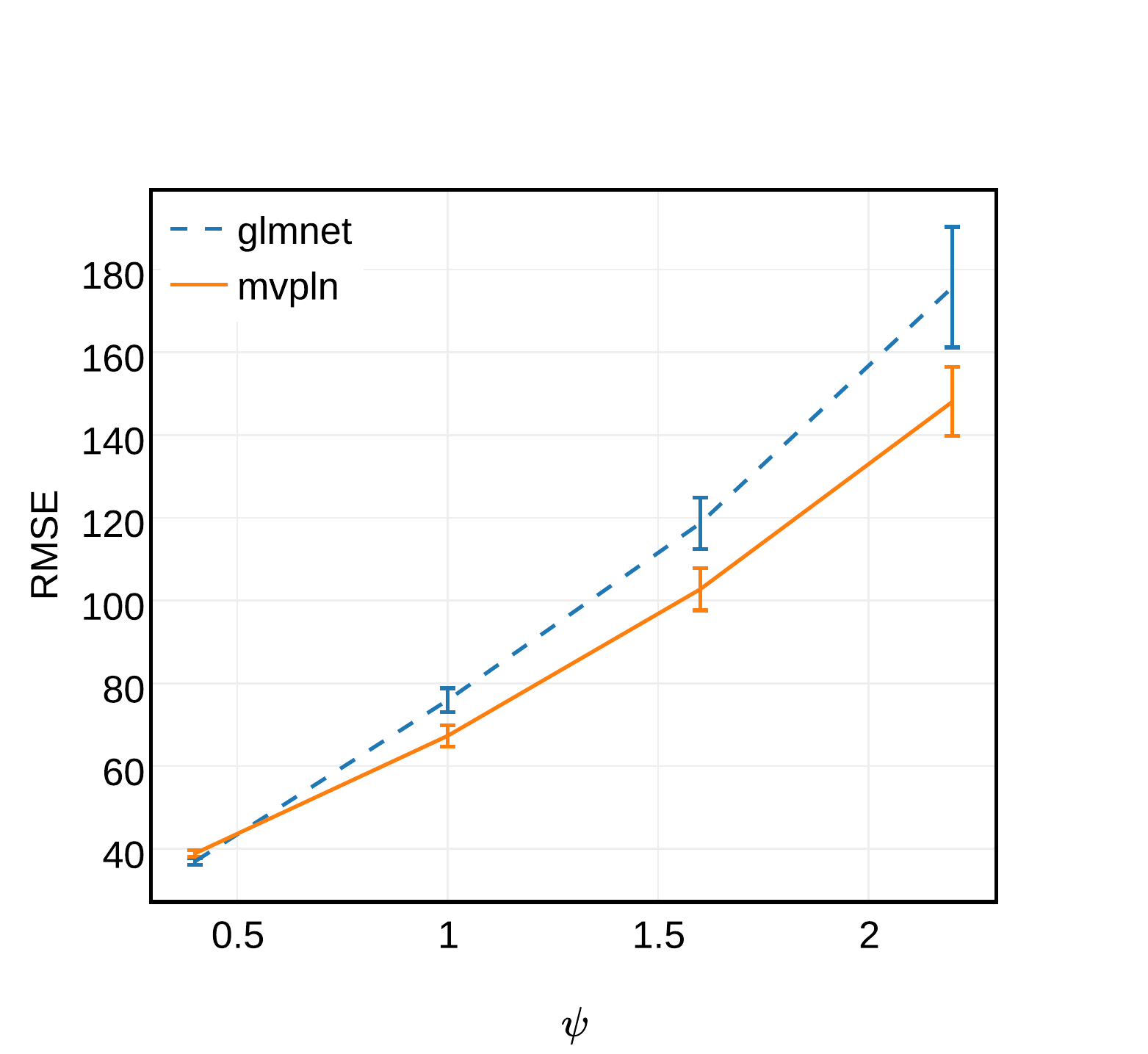}
	\end{minipage}
	\hspace{0.015in}
	\begin{minipage}[t]{0.24\linewidth}
		\centering
		\includegraphics[width=1.6in]{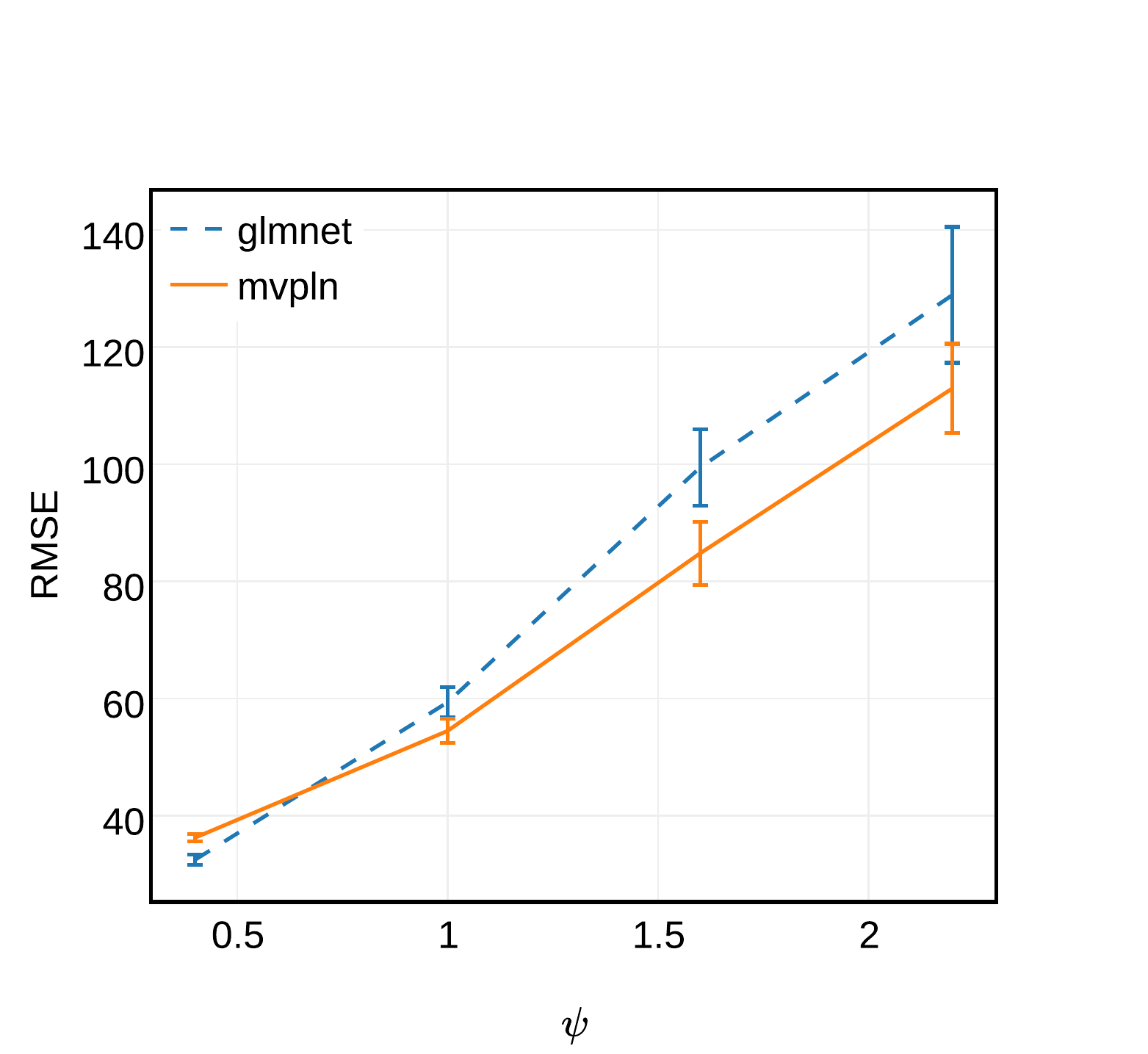}
	\end{minipage}
	\hspace{0.015in}
	\begin{minipage}[t]{0.24\linewidth}
		\centering
		\includegraphics[width=1.6in]{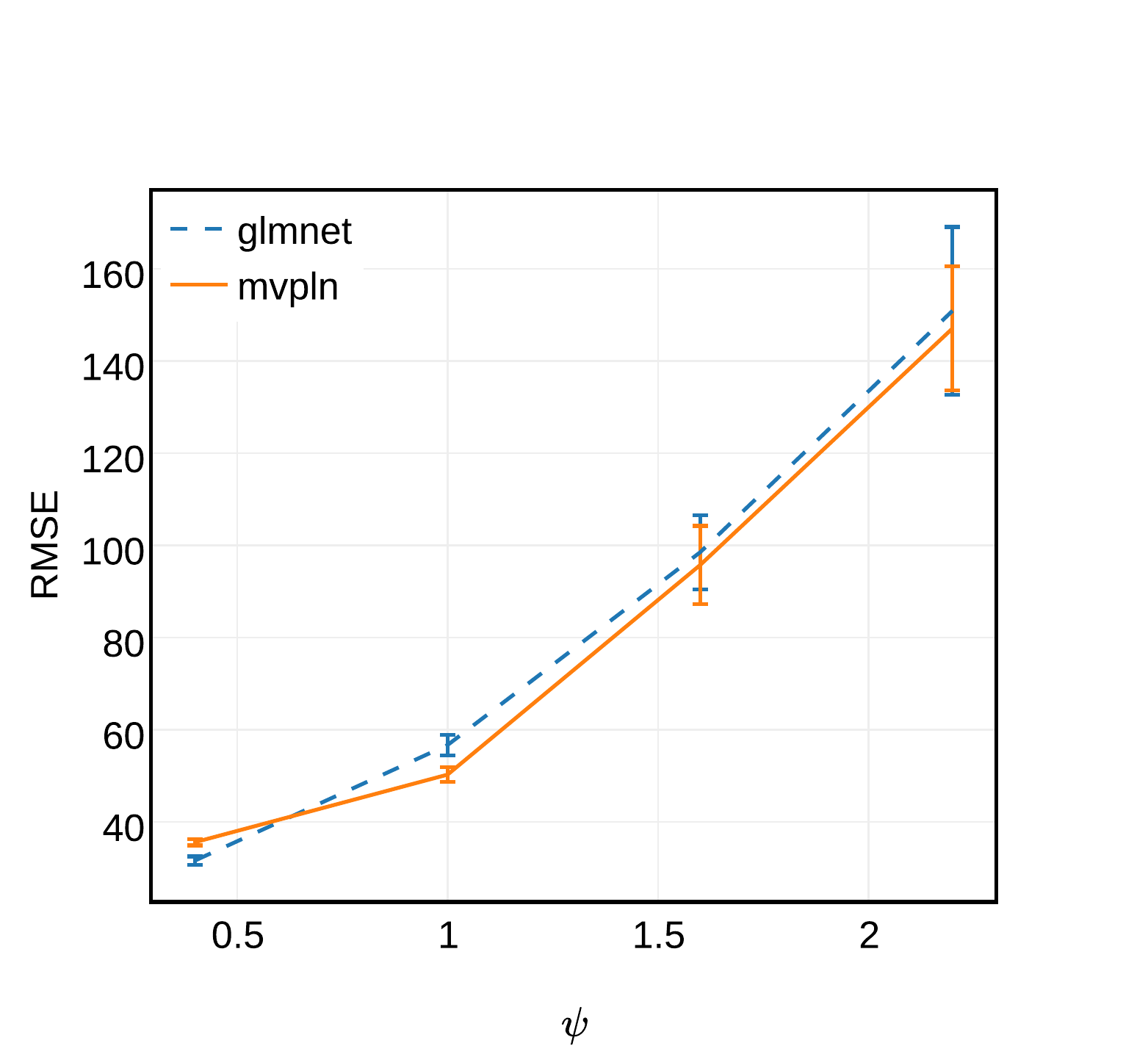}
	\end{minipage}
	\hspace{0.015in}
	\begin{minipage}[t]{0.24\linewidth}
		\centering
		\includegraphics[width=1.6in]{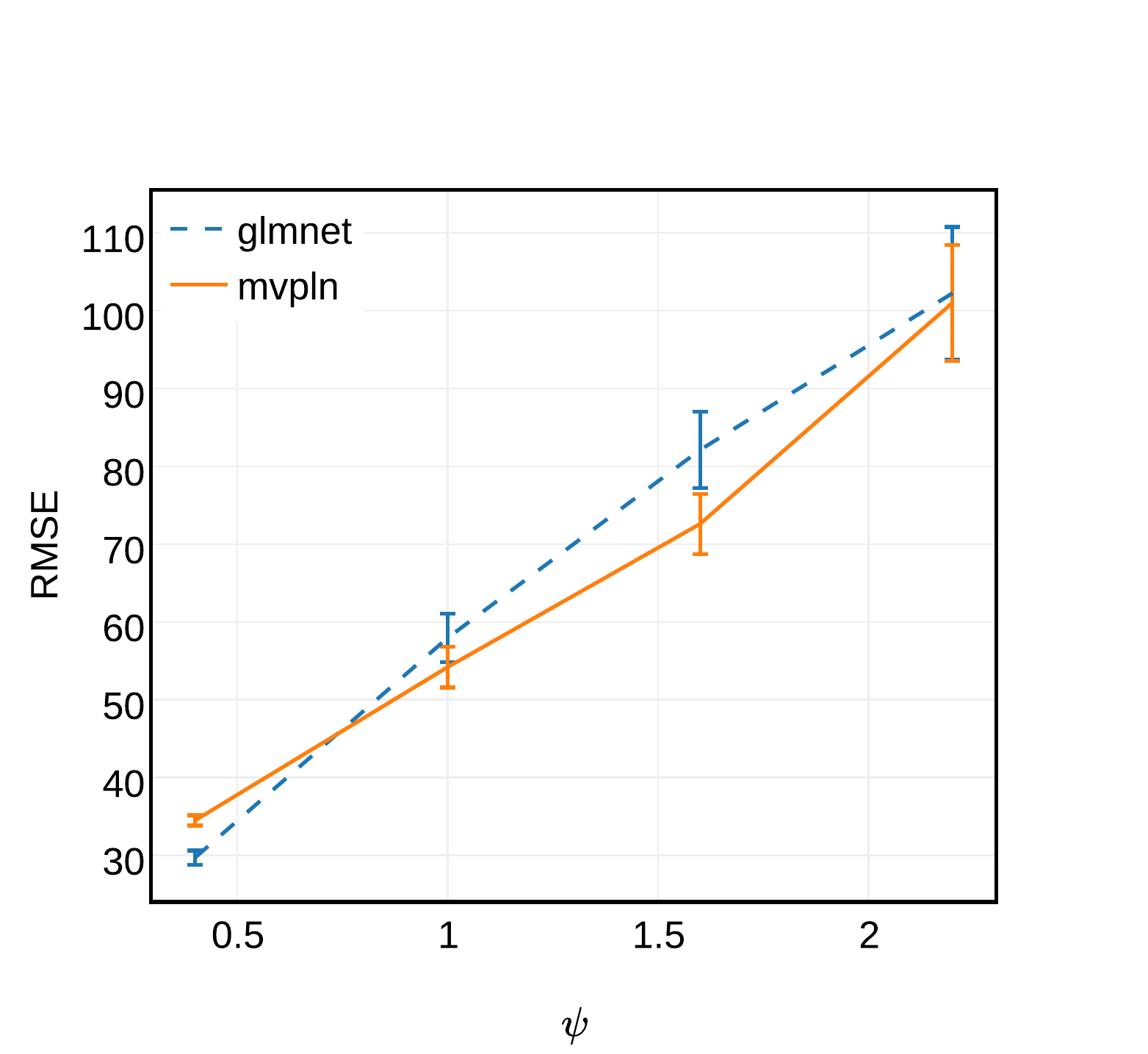}
	\end{minipage}
	\caption{Average rMSE across response dimensions over test data when
		$\boldsymbol{\Omega}$ is random, sparse, banded and diagonal
		(from left to right), and $p \geq n$. The vertical error bars indicate
		the standard deviation.}\label{fig:4.1.2}%\vspace{-0.15in}
\end{figure*}

\subsubsection{Prediction errors}
\label{sec:predict}
To evaluate the prediction performance of the proposed model, we report the
average root-mean-square error (rMSE) across all the response dimensions over
the test data. Figure~\ref{fig:4.1.1} and~\ref{fig:4.1.2} show the average rMSE
for the cases when $p < n$ and $p \geq n$ respectively. These figures show that
when the variations in the simulated data are small ($\psi$ is small), the
prediction performances of the proposed MVPLN model and GLMNET model are
comparable. As the variations in the data increase, the prediction performance
of the proposed MVPLN model becomes better than GLMNET model. This demonstrates
that by incorporating the dependency structures between the count responses,
the proposed MVPLN model improves its prediction performance. However, when the
variations in the data are small, it is difficult for the MVPLN model to take
the advantage of inverse covariance matrix estimation. On the contrary,
approximating the log-likelihood with MCMC techniques would impose negative
effects on the model estimation and prediction accuracy. This is why we observe
that when $\psi$ is small, the proposed MVPLN model sometimes does not perform
as well as the GLMNET model in term of rMSE.

\begin{figure*}[t]
	\begin{minipage}[t]{0.495\linewidth}
		\centering\hspace{-0.15in}
		\includegraphics[width=1.5in]{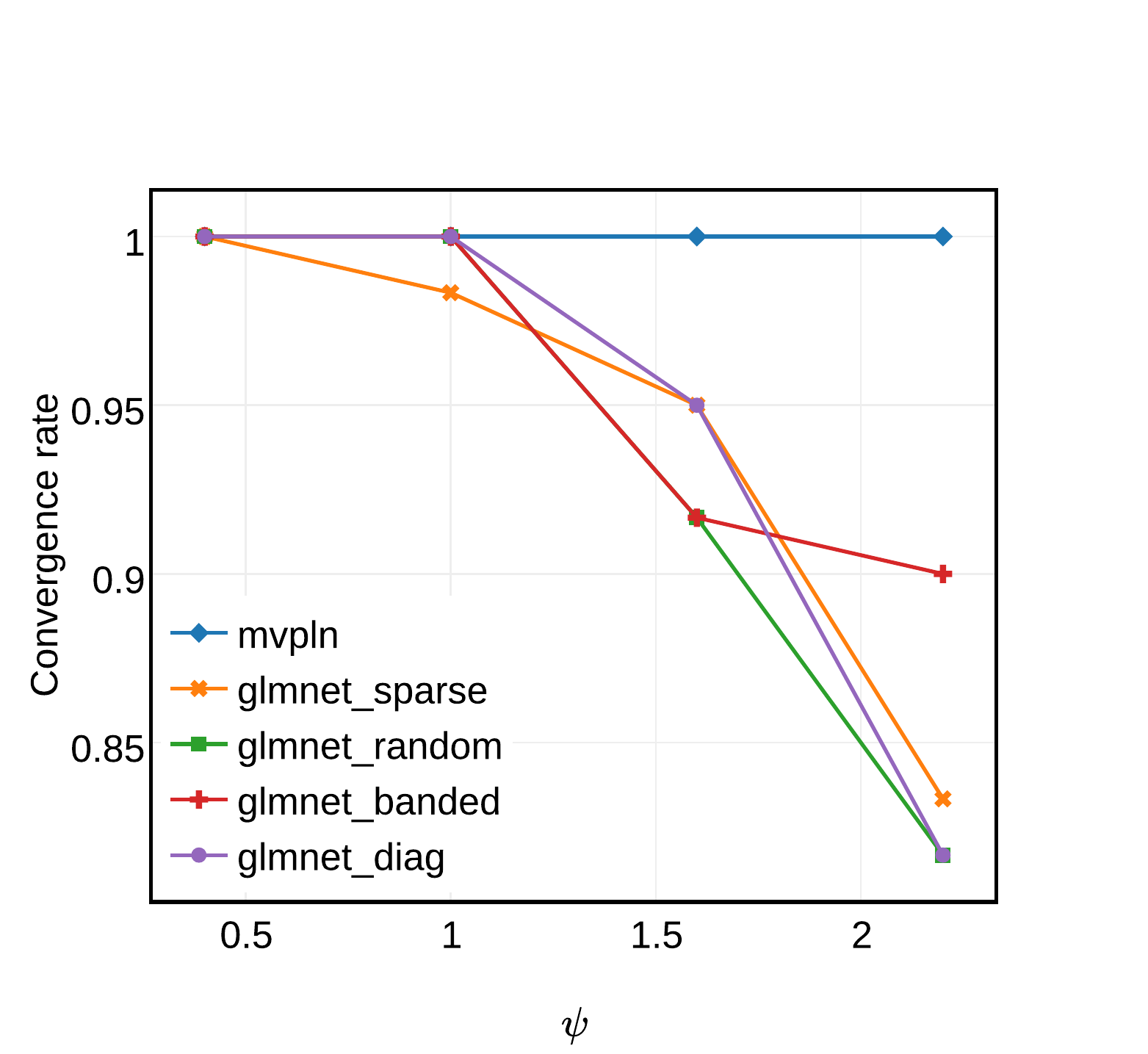}\hspace{-0.2in}
		\includegraphics[width=1.5in]{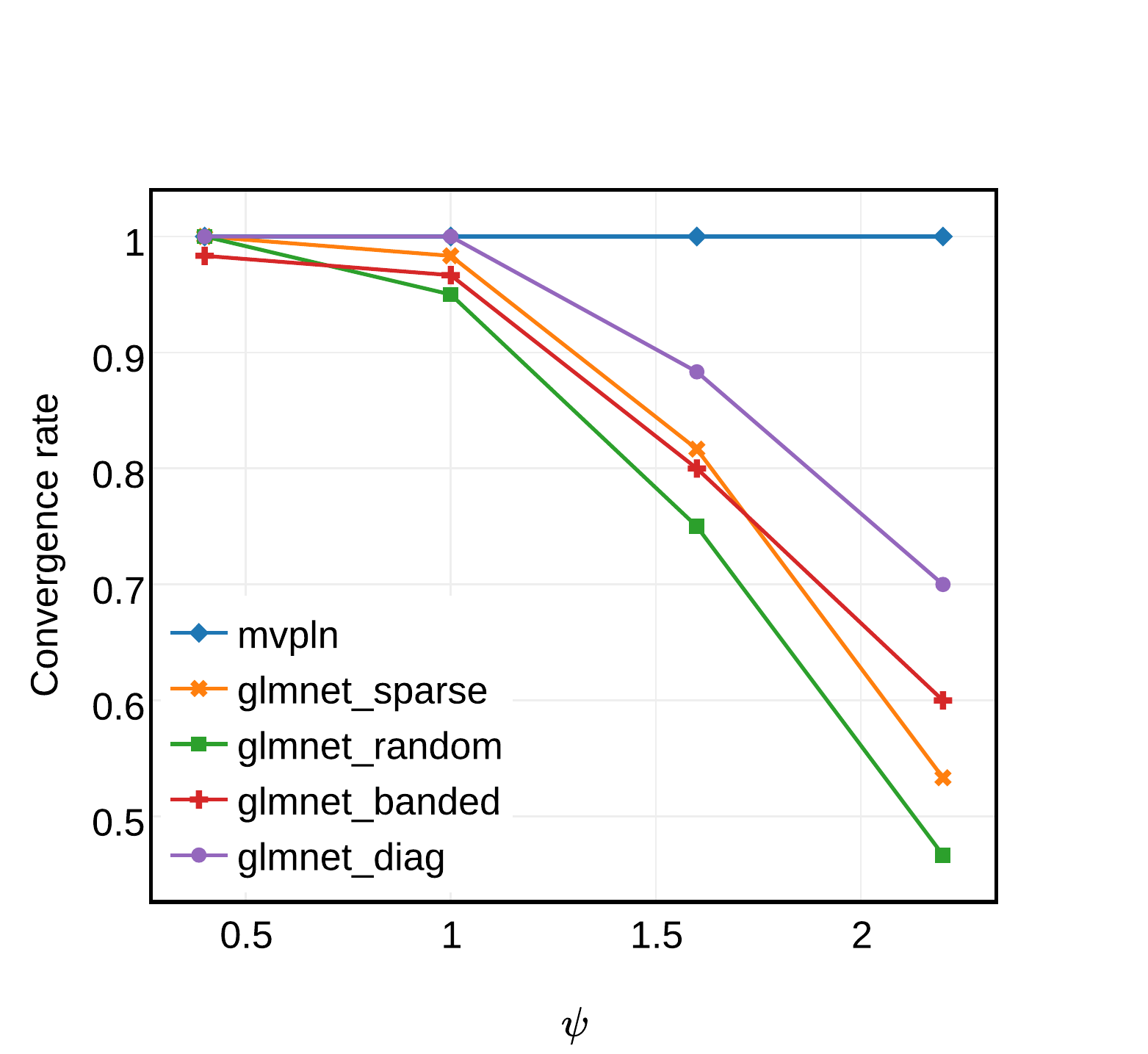}
		\caption{Convergence rate of GLMNET and MVPLN models when $p < n$ (left)
			and $p > n$ (right). Since MVPLN model always converges, we use a
			single line to represent these four scenarios.}\label{fig:conv}
			\vspace{-0.15in}
	\end{minipage}
	\hfill
	\begin{minipage}[t]{0.495\linewidth}
		\centering\hspace{-0.15in}
		\includegraphics[width=1.5in]{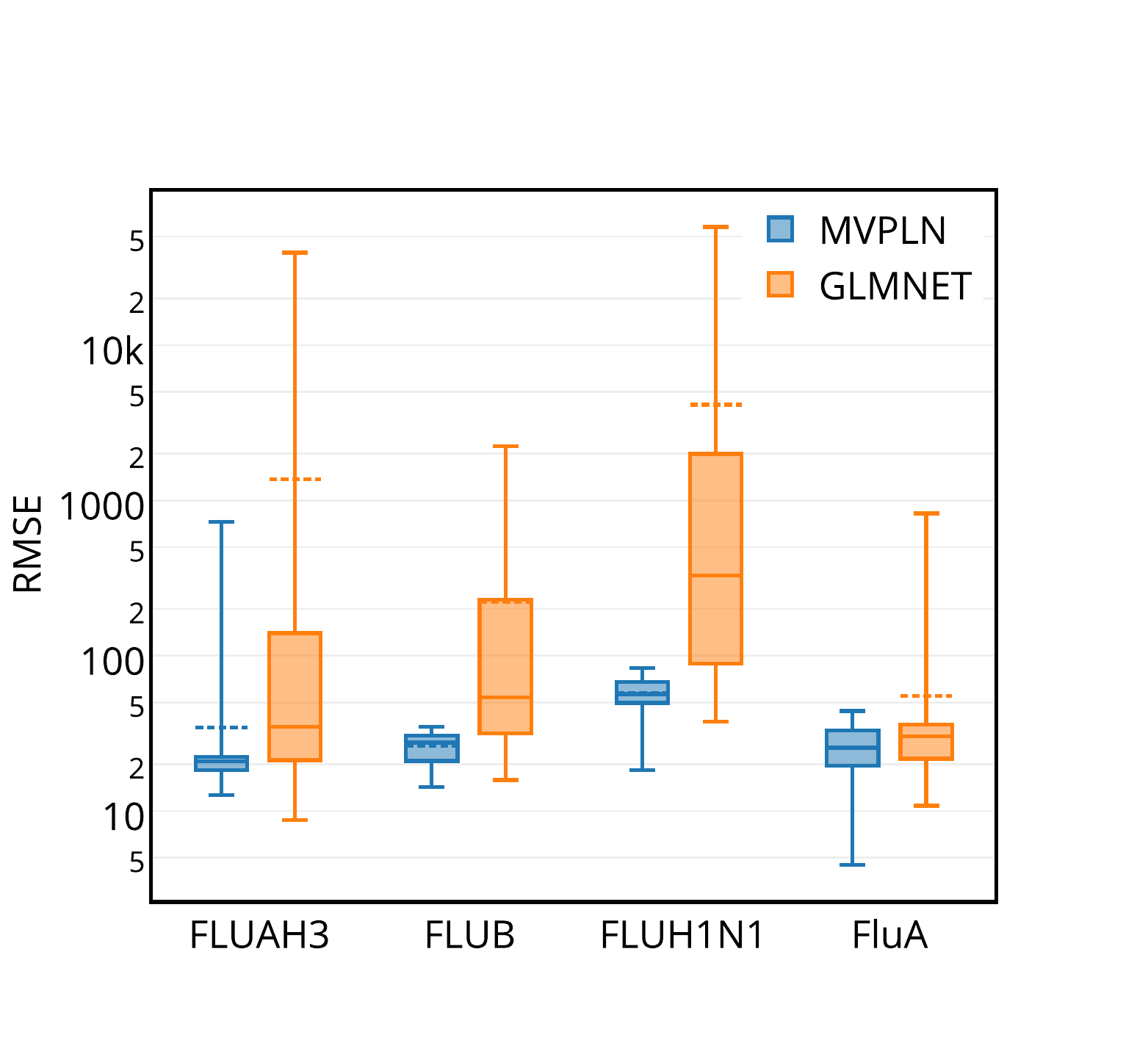}\hspace{-0.2in}
		\includegraphics[width=1.5in]{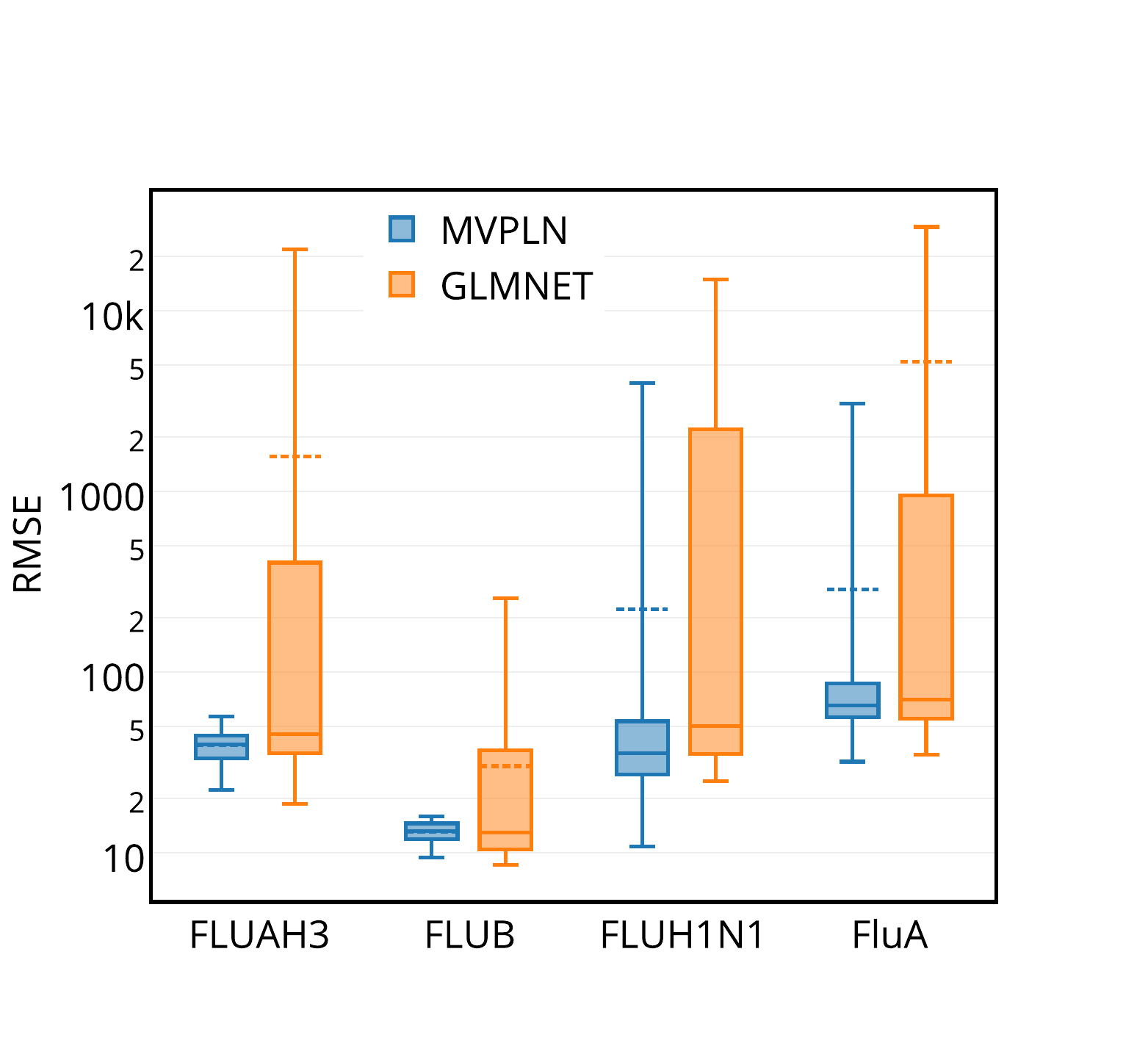}
		\caption{rMSE box plot of MVPLN and GLMNET models on the real ILI dataset
		for the countries of Brazil (left) and Chile (right). The dash lines
		indicate the mean of the rMSE.}\label{fig:box_ili_data}
		%\vspace{-0.15in}
	\end{minipage}
\end{figure*}

\subsubsection{Model convergence}
\label{sec:converge}
Another aspect we would like to emphasize here is the model convergence
performance.  During the experiments over the simulated data, we notice that the
GLMNET model will not always converge in some parameter settings, especially
when $\psi$ is large. As a result, no parameter estimations are given by the
GLMNET model. Figure~\ref{fig:conv} shows the convergence rate (the fraction of
experiment replications that converge and produce valid model estimation) over
the simulated data for various parameter settings. We can see from the figure
that the larger variations (larger $\psi$) in the data, the more frequently the
GLMNET model fails to give a valid model estimation. On the other hand, the
proposed MVPLN model consistently produces the valid model estimation in all of
the scenarios. Such results demonstrate that the proposed MVPLN model is more
robust to the variations in the underlying multivariate data with count
responses.

\subsection{Modeling influenza-like illness case counts}
\label{sec:application}
We apply the proposed MVPLN model to a real influenza-like illness (ILI) dataset
for two Latin American countries, Brazil and Chile, each with four types of ILI
diseases (FLUAH3, FLUB, FLUH1N1 and FLUA). The data were collected from WHO
FluNet~\cite{whoflunet} from May 1st, 2012 to Dec. 27, 2014 ($n = 139$ weeks),
which serves as the multivariate responses of the dataset. The predictors of
this ILI dataset are the weekly counts of $108$ ILI related keywords collected
from the Twitter users of Brazil and Chile during the same period.
% by the EMBERS project~\cite{Ramakrishnan:2014:BNE:2623330.2623373}.
Before applying the proposed MVPLN model, we preprocessed the ILI dataset with
the following approach. We first clustered the $108$ ILI related keywords into
$20$ clusters based on their weekly counts during the selected period using the
k-means algorithm. Then for each cluster, we aggregated the weekly counts
together for the keywords that belong to this cluster, and finally, we scaled
the aggregated keyword counts for each cluster so that it has zero mean and unit
standard deviation. 

%\subsection{Results on the Entire ILI Dataset}
%\label{sec:entire}
It should also be noticed that although this ILI dataset is time-indexed, we
chose to model it as merely a multivariate dataset in our first study here since
the proposed MVPLN model is not specially designed to model time series
datasets. We use $70\%$ of the preprocessed ILI dataset as the training set and
the rest ($30\%$) as the test set. We apply the proposed MVPLN model over the
training set , and compute the rMSE of the test set as the criterion for the
prediction performance of the model. As a comparison, we also apply the GLMNET
model to the same ILI dataset, and compare the rMSE with the proposed MVPLN
model. We repeat this experiment for $60$ independent runs, and for each run, we
shuffle the ILI dataset and re-split the training set and test set.
Figure~\ref{fig:box_ili_data} shows the rMSE box plots of the proposed MVPLN
model and the GLMNET model for Brazil and Chile after removing some extreme
outliers. As we can see from the box plots, although the proposed MVPLN model
generates slightly large rMSE over the test set for some response dimensions
occasionally, in general, the rMSEs of the MVPLN model are much smaller and have
less variation when compared to the GLMNET model for both Brazil and Chile,
which indicates that the proposed MVPLN model is better and more stable in term
of the prediction performance over the real dataset with count responses. Such
results demonstrate that by leveraging the covariance structures between
multiple count responses, the proposed MVPLN model improves the prediction
performance. However, we also notice that for some flu types, the proposed MVPLN
model sometimes generate a large rMSE value, e.g.\ FLUAH3 in the Brazil dataset,
FLUH1N1 and FluA in the Chile dataset. The potential reason for this is likely
that the data shuffling procedure happens to place most of the large-response
data instances into the model training set, which could mislead the model
estimation and result in an overestimation over the test set.

\section{Conclusion}
\label{sec:conclusion}
In this paper, we have proposed and formulated a multivariate Poisson log-normal
model for datasets with count responses. By developing an MCEM algorithm, we
accomplish simultaneous sparse estimations of the regression coefficients and of
the inverse covariance matrix of the model. Results of simulation studies on
synthetic data and an application to a real ILI dataset demonstrate that the
proposed MVPLN model achieves better estimation and prediction performance
versus a classical Lasso regularized Poisson regression model.  Additional
interesting future work for the proposed model are being conducted on the
following lines. (1) Asymptotic properties of the proposed model are being
further investigated; (2) instead of using MCMC techniques, we aim to develop a
better approximation algorithm, e.g.\ using variational
inference~\cite{DBlei:arxiv:2016}; (3) we aim to develop variants of the
proposed model to better deal with count data with over-dispersion and
zero-inflation.

\section*{Appendix}
\appendix
\section{Distribution of multivariate count responses}
Given the multivariate count response $\boldsymbol{\mathcal{Y}}$ and the
predictor $\boldsymbol{x}$, with the conditional independence assumption, the
probability mass function for the multivariate Poisson random variable
$\boldsymbol{\mathcal{Y}}$ is
\begin{align}
	p(\boldsymbol{\mathcal{Y}} = \boldsymbol{y} \mid \boldsymbol{\theta}) = &
	\prod_{i=1}^{q} p(\mathcal{Y}^{(i)} = y^{(i)} \mid \theta^{(i)})
	= \prod_{i=1}^{q} \frac{{\left(\theta^{(i)}\right)}^{y^{(i)}}
	\exp\left(-\theta^{(i)}\right)}{y^{(i)}!}
	\label{eq:3.2.1}
\end{align}
From the specification of the MVPLN model, since $\boldsymbol{\varepsilon} \sim
N(0, \boldsymbol{\Sigma})$, if we let $\boldsymbol{\gamma} =
\boldsymbol{B}^T \boldsymbol{x} + \boldsymbol{\varepsilon}$, we know that
$\boldsymbol{\gamma}$ follows the multivariate normal distribution
$N(\boldsymbol{B}^T \boldsymbol{x}, \boldsymbol{\Sigma})$ with density function:
\begin{align*}
	p(\boldsymbol{\gamma} \mid \boldsymbol{x}) = \frac{1}{{(2 \pi)}^{q/2}
	{|\boldsymbol{\Sigma}|}^{1/2}} \exp \left(-\frac{1}{2}
	{(\boldsymbol{\gamma} - \boldsymbol{B}^{T} \boldsymbol{x})}^T
	\boldsymbol{\Sigma}^{-1} (\boldsymbol{\gamma} - \boldsymbol{B}^T
	\boldsymbol{x})\right)
\end{align*}
Since $\boldsymbol{\theta} = \exp(\boldsymbol{\gamma}) = \exp(\boldsymbol{B}^T
\boldsymbol{x} + \boldsymbol{\varepsilon})$, $\boldsymbol{\theta} \mid
\boldsymbol{x}$ follows the multivariate log-normal distribution, and we can
derive that the density function of $\boldsymbol{\theta} \mid
\boldsymbol{x}$ is:
\begin{align}
	\label{eq:3.2.2}
	p(\boldsymbol{\theta} & \mid \boldsymbol{x}) = p_{\boldsymbol{\gamma}}
	(\log(\boldsymbol{\theta}) \mid \boldsymbol{x})
	\left|\diag\left(\frac{1}{\theta^{(i)}}\right)\right| = \frac{\exp\left(
		-\frac{1}{2} {\left(\log \boldsymbol{\theta} - \boldsymbol{B}^T
		\boldsymbol{x}\right)}^T \boldsymbol{\Sigma}^{-1} \left(\log
		\boldsymbol{\theta} - \boldsymbol{B}^T \boldsymbol{x}\right)\right)
	}{{(2\pi)}^{q/2} |\boldsymbol{\Sigma}|^{1/2} \prod_{i=1}^{q} \theta^{(i)}}.
\end{align}
Thus, the probability mass function for $\boldsymbol{\mathcal{Y}} \mid
\boldsymbol{x}$ is:
\begin{align*}
	p(\boldsymbol{\mathcal{Y}} = \boldsymbol{y} \mid \boldsymbol{x}) = &
		\int_{\boldsymbol{\theta}} p(\boldsymbol{\mathcal{Y}} =
		\boldsymbol{y}, \boldsymbol{\theta} \mid \boldsymbol{x}) d
		\boldsymbol{\theta} = \int_{\boldsymbol{\theta}}
		p(\boldsymbol{\mathcal{Y}} = \boldsymbol{y} \mid \boldsymbol{\theta})
		p(\boldsymbol{\theta} \mid \boldsymbol{x}) d \boldsymbol{\theta} ,
\end{align*}
where $p(\boldsymbol{\mathcal{Y}} = \boldsymbol{y} \mid \boldsymbol{\theta})$
and $p(\boldsymbol{\theta} \mid \boldsymbol{x})$ are specified in
Equation~\eqref{eq:3.2.1} and~\eqref{eq:3.2.2}, respectively.

\section{Monte Carlo E-step in MCEM algorithm}

\subsection{Metropolis Hasting algorithm for sampling $\boldsymbol{\theta}_j$}
Suppose in the MC E-step of iteration $t + 1$, the current estimations of the
model parameters are $\boldsymbol{B}^{(t)}$ and $\boldsymbol{\Sigma}^{(t)}$.
Thus, the conditional distribution of the latent variable $\boldsymbol{\theta}$
given $\boldsymbol{x}, \boldsymbol{y}, \boldsymbol{B}^{(t)}$ and
$\boldsymbol{\Sigma}^{(t)}$ is:
\begin{align}
	p(\boldsymbol{\theta} \mid \boldsymbol{\mathcal{Y}} = \boldsymbol{y},
	\boldsymbol{x}; \boldsymbol{B}^{(t)}, \boldsymbol{\Sigma}^{(t)}) =
	\frac{p(\boldsymbol{\mathcal{Y}} = \boldsymbol{y}, \boldsymbol{\theta} \mid
	\boldsymbol{x}; \boldsymbol{B}^{(t)},
	\boldsymbol{\Sigma}^{(t)})}{p(\boldsymbol{\mathcal{Y}} =
	\boldsymbol{y} \mid \boldsymbol{x}; \boldsymbol{B}^{(t)},
	\boldsymbol{\Sigma}^{(t)})}.
	\label{eq:3.3.2}
\end{align}

\begin{algorithm}[!t]
	\SetAlgoLined
	\SetKwInOut{Input}{input}
	\SetKwInOut{Output}{output}

	\Input{$\boldsymbol{y}_j, \boldsymbol{x}_j, \boldsymbol{B}^{(t)}$ and
	$\boldsymbol{\Sigma}^{(t)}$.}
	\Output{$m$ samples $\boldsymbol{\Theta}_j =
		{\left\{\boldsymbol{\theta}_j^{(1)}, \boldsymbol{\theta}_j^{(2)}, \ldots,
		\boldsymbol{\theta}_j^{(m)}\right\}}^T$.}
	\BlankLine

	Choose $\boldsymbol{\theta}_j^{(0)}$ as initial value, and let $\tau
	\leftarrow 1$\;
	\While{$|\boldsymbol{\Theta}_j| < m$}{
		Draw a candidate $\boldsymbol{\theta}_j^{*}$ from
		$g(\boldsymbol{\theta}_j^{*} \mid \boldsymbol{\theta}_j^{(\tau -
		1)})$\;
		$\alpha \leftarrow
		\min\left(\frac{f(\boldsymbol{\theta}_j^{*}) /
		g(\boldsymbol{\theta}_j^{*} \mid \boldsymbol{\theta}_j^{(\tau -
		1)})}{f(\boldsymbol{\theta}_j^{(\tau - 1)}) /
		g(\boldsymbol{\theta}_j^{(\tau - 1)} \mid
		\boldsymbol{\theta}_j^{*})}, 1 \right)$\;
		Accept $\boldsymbol{\theta}_j^{(*)}$ as
		$\boldsymbol{\theta}_j^{(\tau)}$ with probability $\alpha$\;
		\If{$\boldsymbol{\theta}_j^{(*)}$ is accepted}{
			$\boldsymbol{\Theta}_j \leftarrow \boldsymbol{\Theta}_j \cup
			\{\boldsymbol{\theta}_j^{(\tau)}\}$\;
			$\tau \leftarrow \tau + 1$\;
		}
	}
	\Return{$\boldsymbol{\Theta}_j$}\;

	\caption{Metropolis Hasting algorithm for sampling $\boldsymbol{\theta}_j$}
	\label{alg:A.1}
\end{algorithm}

Then, the expected log-likelihood of the model under $p(\boldsymbol{\theta} \mid
\boldsymbol{\mathcal{Y}} = \boldsymbol{y}, \boldsymbol{x}; \boldsymbol{B}^{(t)},
\boldsymbol{\Sigma}^{(t)})$ would be: 
\begin{align}
	\label{eq:3.3.3}
	Q(\boldsymbol{B}, \boldsymbol{\Sigma} & \mid \boldsymbol{B}^{(t)},
	\boldsymbol{\Sigma}^{(t)}) = E_{p(\boldsymbol{\theta} \mid
	\boldsymbol{\mathcal{Y}} = \boldsymbol{y},
	\boldsymbol{x})}[\mathcal{L}(\boldsymbol{B}, \boldsymbol{\Sigma})] \\
	& = \sum_{j=1}^{n} E_{p(\boldsymbol{\theta}_j \mid \boldsymbol{\mathcal{Y}}
	= \boldsymbol{y}_j, \boldsymbol{x}_j)}[\log p(\boldsymbol{\mathcal{Y}} =
	\boldsymbol{y}_j, \boldsymbol{\theta}_j \mid \boldsymbol{x}_j;
	\boldsymbol{B}, \boldsymbol{\Sigma})]. \nonumber
\end{align}
In order to compute the approximate expected log-likelihood, we adopt the MCMC
technique to sample the $\boldsymbol{\theta}_j$ from $p(\boldsymbol{\theta}_j
\mid \boldsymbol{\mathcal{Y}} = \boldsymbol{y}_j, \boldsymbol{x}_j;
\boldsymbol{B}^{(t)}, \boldsymbol{\Sigma}^{(t)})$. Since $\boldsymbol{y}_j,
\boldsymbol{x}_j, \boldsymbol{B}^{(t)}$ and $\boldsymbol{\Sigma}^{(t)}$ are all
known values, which makes $p(\boldsymbol{\mathcal{Y}} = \boldsymbol{y}_j \mid
\boldsymbol{x}_j; \boldsymbol{B}^{(t)}, \boldsymbol{\Sigma}^{(t)})$ a constant.
In this case, Equation~\eqref{eq:3.3.2} yields
\begin{align*}
	p(\boldsymbol{\theta}_j \mid \boldsymbol{\mathcal{Y}} =
	\boldsymbol{y}_j, & \boldsymbol{x}_j; \boldsymbol{B}^{(t)},
	\boldsymbol{\Sigma}^{(t)})
	\propto p(\boldsymbol{\mathcal{Y}} = \boldsymbol{y}_j,
	\boldsymbol{\theta}_j \mid \boldsymbol{x}_j; \boldsymbol{B}^{(t)},
	\boldsymbol{\Sigma}^{(t)}).
\end{align*}
Let $f(\boldsymbol{\theta}_j) = p(\boldsymbol{\mathcal{Y}} = \boldsymbol{y}_j,
\boldsymbol{\theta}_j \mid \boldsymbol{x}_j; \boldsymbol{B}^{(t)},
\boldsymbol{\Sigma}^{(t)})$ and $g(\boldsymbol{\theta}^{*} \mid
\boldsymbol{\theta})$ be the density function of the proposal distribution.
Algorithm~\ref{alg:A.1} illustrates the Metropolis Hasting algorithm for
sampling $\boldsymbol{\theta}_j$ from $p(\boldsymbol{\theta}_j \mid
\boldsymbol{\mathcal{Y}} = \boldsymbol{y}_j, \boldsymbol{x}_j;
\boldsymbol{B}^{(t)}, \boldsymbol{\Sigma}^{(t)})$.

\subsection{Derivation of the tailored normal distribution as proposal
distribution}
To find the mode of $p(\boldsymbol{\theta}_j \mid \boldsymbol{\mathcal{Y}} =
\boldsymbol{y}_j, \boldsymbol{x}_j; \boldsymbol{B}^{(t)},
\boldsymbol{\Sigma}^{(t)})$, we need to solve the following optimization
problem: 
\begin{align*}
	%\label{eq:3.4.1.2}
	\boldsymbol{\theta}_j^{(0)} = \argmax_{\boldsymbol{\theta}_j} \{\log
		f(\boldsymbol{\theta}_j)\},
\end{align*}
let $F(\boldsymbol{\theta}_j) = \log f(\boldsymbol{\theta}_j) = \log
\left( p(\boldsymbol{\mathcal{Y}} = \boldsymbol{y}_j \mid \boldsymbol{\theta}_j,
\boldsymbol{x}_j; \boldsymbol{B}^{(t)}, \boldsymbol{\Sigma}^{(t)})
p(\boldsymbol{\theta}_j \mid \boldsymbol{x}_j; \boldsymbol{B}^{(t)},
\boldsymbol{\Sigma}^{(t)})\right)$. By combining
Equation~\eqref{eq:3.2.1} and~\eqref{eq:3.2.2} together, we can derive that:
\begin{align}
	F(\boldsymbol{\theta}_j) & = {(\boldsymbol{y}_j - \boldsymbol{1})}^T \log
	\boldsymbol{\theta}_j - \boldsymbol{1}^T \boldsymbol{\theta}_j - \frac{1}{2}
	{\left( \log \boldsymbol{\theta}_j - {\boldsymbol{B}^{(t)}}^T
	\boldsymbol{x}_j \right)}^T {\boldsymbol{\Sigma}^{(t)}}^{-1} \left( \log
	\boldsymbol{\theta}_j - {\boldsymbol{B}^{(t)}}^T \boldsymbol{x}_j \right) +
	C,
	\label{eq:3.4.1.3}
\end{align}
where $\boldsymbol{1}$ denotes a column vector of $1$s, and $C$ represents the
sum of all the constants in $\log f(\boldsymbol{\theta}_j)$. Then, the first
order and second order derivatives of $F(\boldsymbol{\theta}_j)$ w.r.t.\
$\boldsymbol{\theta}_j$ are
\begin{align}
	\nabla F(\boldsymbol{\theta}_j) = & \frac{d F(\boldsymbol{\theta}_j)}{d
		\boldsymbol{\theta}_j} = \diag \left(\frac{1}{\theta_j^{(i)}}\right)
		\left[(\boldsymbol{y}_j - \boldsymbol{1}) -
			{\boldsymbol{\Sigma}^{(t)}}^{-1} \left(\log \boldsymbol{\theta}_j -
	{\boldsymbol{B}^{(t)}}^{T} \boldsymbol{x}_j \right)\right] - \boldsymbol{1}
	\label{eq:3.4.1.4} \\
	\boldsymbol{H}(\boldsymbol{\theta}_j) = & \diag\left(-\frac{y_j^{(i)} -
		1}{{\theta_j^{(i)}}^2}\right) +
		\diag\left( -\frac{1}{{\theta_j^{(i)}}^2}\right)
		\diag\left({\boldsymbol{\Sigma}^{(t)}}^{-1} \left(\log
		\boldsymbol{\theta}_j - {\boldsymbol{B}^{(t)}}^T \boldsymbol{x}_j
		\right)\right) \nonumber \\
		& + \diag\left(\frac{1}{\theta_j^{(i)}}\right)
		{\boldsymbol{\Sigma}^{(t)}}^{-1}
		\diag\left(\frac{1}{\theta_j^{(i)}}\right) \label{eq:3.4.1.5}
\end{align}
Let $\nabla F(\boldsymbol{\theta}_j) = 0$, and we could get that the initial
value $\boldsymbol{\theta}_j^{(0)}$ of the location parameter for the tailored
normal distribution is the solution to the following equation:
\begin{align}
	\boldsymbol{\theta}_j + {\boldsymbol{\Sigma}^{(t)}}^{-1} \log
	\boldsymbol{\theta}_j = \boldsymbol{y}_j - \boldsymbol{1} +
	{\boldsymbol{\Sigma}^{(t)}}^{-1} {\boldsymbol{B}^{(t)}}^T
	\boldsymbol{x}_j
	\label{eq:3.4.1.6}
\end{align}
which can be solved by any numerical root discovering algorithms. However,
taking performance issue into account, we let $\boldsymbol{\kappa}_j = \log
\boldsymbol{\theta}_j$, and adopt a linear approximation to
$\boldsymbol{e}^{\boldsymbol{\kappa}_j}$ with its first order Taylor expansion
at $\boldsymbol{\kappa}_j^{(0)} = \log \boldsymbol{y}_j$. In this case,
Equation~\eqref{eq:3.4.1.6} becomes:
\begin{align}
	\boldsymbol{e}^{\boldsymbol{\kappa}_j^{(0)}} +
	\diag \left(\boldsymbol{e}^{\boldsymbol{\kappa}_j^{(0)}}\right) 
	\left(\boldsymbol{\kappa}_j - \boldsymbol{\kappa}_j^{(0)}\right) +
	{\boldsymbol{\Sigma}^{(t)}}^{-1} \boldsymbol{\kappa}_j = \boldsymbol{y}_j -
	\boldsymbol{1} + {\boldsymbol{\Sigma}^{(t)}}^{-1} {\boldsymbol{B}^{(t)}}^T
	\boldsymbol{x}_j.
	\label{eq:B.2.1}
\end{align}
Solving Equation~\eqref{eq:B.2.1} for $\boldsymbol{\kappa}_j$, the location
parameter (mean) $\boldsymbol{\theta}_j$ of the tailored normal distribution is
given by  $\boldsymbol{\theta}_j^{(0)} = \boldsymbol{e}^{\boldsymbol{\kappa}_j}$
where
\begin{align*}
	\boldsymbol{\kappa}_j = &
	{\left( \diag \left(\boldsymbol{e}^{\boldsymbol{\kappa}_j^{(0)}}\right) +
	{\boldsymbol{\Sigma}^{(t)}}^{-1}\right)}^{-1} \Big( \boldsymbol{y}_j -
	\boldsymbol{1} + {\boldsymbol{\Sigma}^{(t)}}^{-1} {\boldsymbol{B}^{(t)}}^T
	\boldsymbol{x}_j +
	\diag\left(\boldsymbol{e}^{\boldsymbol{\kappa}_j^{(0)}}\right)
	\boldsymbol{\kappa}_j^{(0)} -
	\boldsymbol{e}^{\boldsymbol{\kappa}_j^{(0)}}\Big),
\end{align*}
and the covariance matrix is given by $\tau
(-\boldsymbol{H}(\boldsymbol{\theta}_j^{(0)}))^{-1}$. In the case that the
covariance matrix $\tau (-\boldsymbol{H}(\boldsymbol{\theta}_j^{(0)}))^{-1}$ is
not positive semidefinite, the nearest positive semidefinite matrix to $\tau
(-\boldsymbol{H}(\boldsymbol{\theta}_j^{(0)}))^{-1}$ is used to replace $\tau
(-\boldsymbol{H}(\boldsymbol{\theta}_j^{(0)}))^{-1}$.

\section{M-step in the MCEM algorithm}
\subsection{The optimization problem in M-step}
The joint distribution of $(\boldsymbol{\mathcal{Y}} = \boldsymbol{y}_j,
\boldsymbol{\theta}_j^{(\tau)})$ given $\boldsymbol{x}_j, \boldsymbol{B}^{(t)}$
and $\boldsymbol{\Sigma}^{(t)}$ is:
\begin{align*}
	p(\boldsymbol{\mathcal{Y}} = \boldsymbol{y}_j,
	\boldsymbol{\theta}_j^{(\tau)} \mid \boldsymbol{x}_j; \boldsymbol{B}^{(t)},
	\boldsymbol{\Sigma}^{(t)}) = p (\boldsymbol{\mathcal{Y}} =
	\boldsymbol{y}_j \mid \boldsymbol{\theta}_j^{(\tau)})
	p(\boldsymbol{\theta}_j^{(\tau)} \mid \boldsymbol{x}_j;
	\boldsymbol{B}^{(t)}, \boldsymbol{\Sigma}^{(t)})
\end{align*}
where $p (\boldsymbol{\mathcal{Y}} = \boldsymbol{y}_j \mid
\boldsymbol{\theta}_j^{(\tau)})$ and $p(\boldsymbol{\theta}_j^{(\tau)} \mid
\boldsymbol{x}_j; \boldsymbol{B}^{(t)}, \boldsymbol{\Sigma}^{(t)})$ are given by
Equation~\eqref{eq:3.2.1} and~\eqref{eq:3.2.2} respectively. Let
$\boldsymbol{\Omega} = \boldsymbol{\Sigma}^{-1}$ and
$\boldsymbol{\varphi}_{\tau, j} = (\log \boldsymbol{\theta}_j^{(\tau)} -
\boldsymbol{B}^T \boldsymbol{x}_j)$. Combining the approximated expected
log-likelihood we derived in the MC E-step in Section 2.2.1 (Equation (7) in the
paper), we can reformulate $\tilde{Q}(\boldsymbol{B}, \boldsymbol{\Sigma} \mid
\boldsymbol{B}^{(t)}, \boldsymbol{\Sigma}^{(t)})$ as:
\begin{align}
	\tilde{Q}( \boldsymbol{B}, \boldsymbol{\Sigma} \mid \boldsymbol{B}^{(t)},
	\boldsymbol{\Sigma}^{(t)}) & = - \frac{1}{n} \sum_{j=1}^{n} \frac{1}{m}
	\sum_{\tau = 1}^{m} \bigg[ {\left(\log \boldsymbol{\theta}_j^{(\tau)} -
	\boldsymbol{B}^T \boldsymbol{x}_j \right)}^T \boldsymbol{\Omega} \left(\log
	\boldsymbol{\theta}_j^{(\tau)} - \boldsymbol{B}^T \boldsymbol{x}_j \right) -
	\log |\boldsymbol{\Omega} | \bigg] \nonumber \\
	& = -\frac{1}{mn} \tr\left(\boldsymbol{\Phi}^T \boldsymbol{\Phi}
	\boldsymbol{\Omega}\right) + \log |\boldsymbol{\Omega}|.
	\label{eq:3.4.2.2}
\end{align}
Then, the optimization problem we need to solve in the M-step is:
\begin{align}
	\boldsymbol{B}^{(t+1)}, \boldsymbol{\Sigma}^{(t+1)} & =
	\argmin_{\boldsymbol{B}, \boldsymbol{\Sigma}}
	\left\{-\tilde{Q}(\boldsymbol{B}, \boldsymbol{\Sigma} \mid
	\boldsymbol{B}^{(t)}, \boldsymbol{\Sigma}^{(t)}) + \lambda_1
	||\boldsymbol{B}||_{1} + \lambda_2 ||\boldsymbol{\Sigma}^{-1}||_{1} \right\}
	\nonumber \\
	& = \argmin_{\boldsymbol{B}, \boldsymbol{\Omega}}\left\{\frac{1}{mn}
	\tr\left(\boldsymbol{\Phi}^T \boldsymbol{\Phi} \boldsymbol{\Omega}\right) - 
	\log |\boldsymbol{\Omega}| + \lambda_1 ||\boldsymbol{B}||_{1} + \lambda_2
	||\boldsymbol{\Omega}||_{1}\right\}
	\label{eq:3.4.2.1}
\end{align}

\subsection{Approach to solve $\boldsymbol{B}$ approximately when
$\boldsymbol{\Omega}$ fixed}
When $\boldsymbol{\Omega}$ is fixed at $\boldsymbol{\Omega}_0$, we have the
following convex optimization problem:
\begin{align}
	\boldsymbol{B}(\boldsymbol{\Omega}_0) = \argmin_{\boldsymbol{B}}
	\bigg\{ \frac{1}{mn} \tr\left( \boldsymbol{\Phi}^T \boldsymbol{\Phi}
	\boldsymbol{\Omega}_0 \right) + \lambda_1 ||\boldsymbol{B}||_{1} \bigg\}.
	\label{eq:C.2.1}
\end{align}
The $l_{1}$ matrix norm penalty in Equation~\eqref{eq:C.2.1} could be
approximated with the following approach
\begin{align*}
	||\boldsymbol{B}||_{1} \approx \tr\left({\boldsymbol{B}'}^T \boldsymbol{B}'
	\right), \quad \text{where}~\boldsymbol{B}' = \boldsymbol{B} \circ
	\frac{1}{\sqrt{|\hat{\boldsymbol{B}}|}}.
\end{align*}
Here, $\circ$ denotes the Hadamard (element-wise) product. If we write
$\boldsymbol{\Phi}$ into the following block matrix
\begin{align*}
	\boldsymbol{\Phi} = \left[
		\begin{array}{c}
			\log \boldsymbol{\Theta}_1 - \boldsymbol{X}_1 \boldsymbol{B} \\
			\log \boldsymbol{\Theta}_2 - \boldsymbol{X}_2 \boldsymbol{B} \\
			\vdots \\
			\log \boldsymbol{\Theta}_n - \boldsymbol{X}_n \boldsymbol{B} \\
		\end{array}\right],
\end{align*}
where $\boldsymbol{X}_j$ is $m \times p$ matrix with each row being
$\boldsymbol{x}_j$ for all $j = 1, 2, \ldots, n$, the objective function of the
optimization problem in~\eqref{eq:C.2.1} can be written as:
\begin{align}
	\eta(&\boldsymbol{B}) = \lambda_{1} \tr\left( {\boldsymbol{B}'}^T
	\boldsymbol{B}' \right) + \frac{1}{mn} \sum_{j=1}^{n} \tr\Big({(\log
	\boldsymbol{\Theta}_j - \boldsymbol{X}_j \boldsymbol{B})}^T (\log
	\boldsymbol{\Theta}_j - \boldsymbol{X}_j \boldsymbol{B})
	\boldsymbol{\Omega}_0 \Big).
	\label{eq:3.4.2.6}
\end{align}
Taking the first order derivative of $\eta(\boldsymbol{B})$ w.r.t.\
$\boldsymbol{B}$ and setting it to zero, we have
\begin{align}
	\left(\sum_{j=1}^{n} \boldsymbol{X}_{j}^T \boldsymbol{X}_j \right)
	\boldsymbol{B} \boldsymbol{\Omega}_0 + \boldsymbol{B} \circ
	\frac{\lambda_1 m n}{|\hat{\boldsymbol{B}}|} 
	 = \left(\sum_{j=1}^{n} \boldsymbol{X}_{j}^T (\log
	\boldsymbol{\Theta}_j) \right) \boldsymbol{\Omega}_0
	\label{eq:3.4.2.7}
\end{align}

If we let $\Big(\sum\limits_{j=1}^{n} \boldsymbol{X}_j^T (\log
\boldsymbol{\Theta}_j)\Big) \boldsymbol{\Omega}_0 = \boldsymbol{H}$ and
$\sum\limits_{j = 1}^{n} \boldsymbol{X}_j^T \boldsymbol{X}_j = \boldsymbol{S}$,
and apply the matrix vectorization operator $\vectorize(\cdot)$ to both sides of
Equation~\eqref{eq:3.4.2.7}, we have:
\begin{align*}	
	\left( \boldsymbol{\Omega}_0^{T} \otimes \boldsymbol{S} \right)
	\vectorize\left(\boldsymbol{B}\right) + \vectorize\left(\frac{\lambda_1 m
	n}{|\hat{\boldsymbol{B}|}}\right) \circ
	\vectorize\left(\boldsymbol{B}\right) = \vectorize\left(
	\boldsymbol{H} \right).
\end{align*}
Here, $\otimes$ represents the Kronecker product. By pulling
$\vectorize(\boldsymbol{B})$ out from the left hand side of the above equation,
we can get: 
\begin{align*}
	\left[\boldsymbol{\Omega}_0 \otimes \boldsymbol{S} + \diag\left(
	\vectorize\left( \frac{\lambda_1 m
	n}{|\boldsymbol{B}_{\mathit{est}}|}\right)\right)\right]
	\vectorize(\boldsymbol{B}) = \vectorize(\boldsymbol{H}).
\end{align*}
Thus, the solution to the optimization problem in Equation~\eqref{eq:C.2.1} is
\begin{align}
	\vectorize(\boldsymbol{B}) = {\left[\boldsymbol{\Omega}_0 \otimes
	\boldsymbol{S} + \diag\left(\vectorize\left(\frac{\lambda_1 m
	n}{|\hat{\boldsymbol{B}}|}\right)\right)\right]}^{-1} \hspace{-0.2cm}
	\vectorize(\boldsymbol{H}),
	\label{eq:C.2.4}
\end{align}
and the estimated coefficient matrix $\boldsymbol{B}$ can be obtained by
reorganizing the $\vectorize(\boldsymbol{B})$ in the above equation.

\subsection{Algorithm pseudo code for M-step}
By solving $\boldsymbol{B}$ and $\boldsymbol{\Omega}$ alternatively with the
other fixed at the value of the last estimate until convergence, we will obtain
the MLE of the coefficient matrix $\boldsymbol{B}$ and inverse covariance matrix
$\boldsymbol{\Omega}$ for the current iteration of MCEM algorithm.
Algorithm~\ref{alg:3.4.2.1} summarizes the M-step of the MCEM algorithm.
\begin{algorithm}[t]
	\SetAlgoLined
	\SetKwInOut{Input}{input}
	\SetKwInOut{Output}{output}
	\SetKwFunction{KwFnGLasso}{Graphical\_Lasso}

	\Input{$\boldsymbol{X}, \{\boldsymbol{\Theta}_j\},
	\boldsymbol{\Omega}_0, \boldsymbol{B}_0, \lambda_1$ and $\lambda_2$.}
	\Output{MLE of $\boldsymbol{B}$ and $\boldsymbol{\Omega}$.}
	\BlankLine

	$t \leftarrow -1$\;
	\Repeat{convergence}{
		$t \leftarrow t + 1$\;
		$\boldsymbol{\Phi} \leftarrow \left[\begin{array}{c}
			\log \boldsymbol{\Theta}_1 - \boldsymbol{X}_1
			\boldsymbol{B}^{(t)} \\
			\log \boldsymbol{\Theta}_2 - \boldsymbol{X}_2
			\boldsymbol{B}^{(t)} \\
			\vdots \\
			\log \boldsymbol{\Theta}_n - \boldsymbol{X}_n
			\boldsymbol{B}^{(t)} \\
		\end{array}\right]$\;
		$\boldsymbol{\Omega}^{(t+1)} \leftarrow$
		\KwFnGLasso{$\boldsymbol{\Phi}, \lambda_2$}\;
		$\boldsymbol{S} \leftarrow \sum_{j=1}^{n} \boldsymbol{X}_j^T
		\boldsymbol{X}_j$\;
		$\boldsymbol{H} \leftarrow \sum_{j=1}^{n} \boldsymbol{X}_j^T (\log
		\boldsymbol{\Theta}_j) \boldsymbol{\Omega}^{(t+1)}$\;
		\vspace{-0.2cm}
		{\setlength{\abovedisplayskip}{-7pt} \setlength{\belowdisplayskip}{0pt}
		\begin{flalign*}
			\boldsymbol{B}^{(t+1)} \leftarrow \bigg[
				\boldsymbol{\Omega}^{(t+1)} \otimes \boldsymbol{S} +
				\diag\left(\vectorize\left(\frac{\lambda_1 m
				n}{|\boldsymbol{B}^{(t)}|}\right)\right)\bigg]^{-1}
			\vectorize(\boldsymbol{H});&&
		\end{flalign*}}
	}
	\Return{($\boldsymbol{B}^{(t+1)}, \boldsymbol{\Omega}^{(t+1)})$}\;

	\caption{M-step of the MCEM algorithm}
	\label{alg:3.4.2.1}
\end{algorithm}

\bibliographystyle{abbrvnat}
\bibliography{reference}

\begin{thebibliography}{31}
\providecommand{\natexlab}[1]{#1}
\providecommand{\url}[1]{\texttt{#1}}
\expandafter\ifx\csname urlstyle\endcsname\relax
  \providecommand{\doi}[1]{doi: #1}\else
  \providecommand{\doi}{doi: \begingroup \urlstyle{rm}\Url}\fi

\bibitem[who(2015)]{whoflunet}
{WHO FluNet}, 2015.
\newblock \url{http://www.who.int/influenza/gisrs_laboratory/flunet/en/}.

\bibitem[Argyriou et~al.(2007{\natexlab{a}})Argyriou, Evgeniou, and
  Pontil]{NIPS2006_3143}
A.~Argyriou, T.~Evgeniou, and M.~Pontil.
\newblock Multi-task feature learning.
\newblock In \emph{NIPS}, pages 41--48, 2007{\natexlab{a}}.

\bibitem[Argyriou et~al.(2007{\natexlab{b}})Argyriou, Micchelli, Pontil, and
  Ying]{conf:nips:ArgyriouMPY07}
A.~Argyriou, C.~A. Micchelli, M.~Pontil, and Y.~Ying.
\newblock A spectral regularization framework for multi-task structure
  learning.
\newblock In \emph{NIPS}, 2007{\natexlab{b}}.

\bibitem[Blei et~al.(2016)Blei, Kucubelbir, and McAuliffe]{DBlei:arxiv:2016}
D.~M. Blei, A.~Kucubelbir, and J.~D. McAuliffe.
\newblock Variational inference: A review for statisticians, 2016.
\newblock \url{https://arxiv.org/abs/1601.00670}.

\bibitem[Chen and Chen(2008)]{Jiahua:v:95:y:2008:i:3}
J.~Chen and Z.~Chen.
\newblock {Extended Bayesian information criteria for model selection with
  large model spaces}.
\newblock \emph{Biometrika}, 95\penalty0 (3):\penalty0 759--771, 2008.

\bibitem[Chen et~al.(2011)Chen, Zhou, and Ye]{Chen:2011:ILG:2020408.2020423}
J.~Chen, J.~Zhou, and J.~Ye.
\newblock Integrating low-rank and group-sparse structures for robust
  multi-task learning.
\newblock In \emph{KDD '11}, pages 42--50, 2011.

\bibitem[Chib et~al.(1998)Chib, Greenberg, and Winkelmann]{Chib199833}
S.~Chib, E.~Greenberg, and R.~Winkelmann.
\newblock Posterior simulation and bayes factors in panel count data models.
\newblock \emph{Journal of Econometrics}, 86\penalty0 (1):\penalty0 33 -- 54,
  1998.

\bibitem[El-Basyouny and Sayed(2009)]{ElBasyouny2009820}
K.~El-Basyouny and T.~Sayed.
\newblock Collision prediction models using multivariate poisson-lognormal
  regression.
\newblock \emph{Accident Analysis and Prevention}, 41\penalty0 (4):\penalty0
  820 -- 828, 2009.

\bibitem[Foygel and Drton(2010)]{NIPS2010_4087}
R.~Foygel and M.~Drton.
\newblock Extended bayesian information criteria for gaussian graphical models.
\newblock In \emph{NIPS}, pages 604--612, 2010.

\bibitem[Friedman et~al.(2007)Friedman, Hastie, H{\"o}fling, and
  Tibshirani]{friedman2007:pathwise}
J.~Friedman, T.~Hastie, H.~H{\"o}fling, and R.~Tibshirani.
\newblock {Pathwise coordinate optimization}.
\newblock \emph{The Annals of Applied Statistics}, 1\penalty0 (2):\penalty0
  302--332, 2007.

\bibitem[Friedman et~al.(2008)Friedman, Hastie, and
  Tibshirani]{Friedman:2134265}
J.~Friedman, T.~Hastie, and R.~Tibshirani.
\newblock {Sparse inverse covariance estimation with the graphical lasso}.
\newblock \emph{Biostatistics}, 9\penalty0 (3):\penalty0 432--441, July 2008.

\bibitem[Friedman et~al.(2014)Friedman, Hastie, Simon, and
  Tibshirani]{r:glmnet}
J.~Friedman, T.~Hastie, N.~Simon, and R.~Tibshirani.
\newblock Lasso and elastic-net regularized generalized linear models, 2014.
\newblock URL \url{http://cran.r-project.org/web/packages/glmnet/glmnet.pdf}.

\bibitem[Gong et~al.(2012{\natexlab{a}})Gong, Ye, and shui
  Zhang]{NIPS2012_4729}
P.~Gong, J.~Ye, and C.~shui Zhang.
\newblock Multi-stage multi-task feature learning.
\newblock In \emph{NIPS}, 2012{\natexlab{a}}.

\bibitem[Gong et~al.(2012{\natexlab{b}})Gong, Ye, and
  Zhang]{Gong:2012:RMF:2339530.2339672}
P.~Gong, J.~Ye, and C.~Zhang.
\newblock Robust multi-task feature learning.
\newblock In \emph{KDD '12}, pages 895--903, 2012{\natexlab{b}}.

\bibitem[Gong et~al.(2014)Gong, Zhou, Fan, and
  Ye]{Gong:2014:EMF:2623330.2623641}
P.~Gong, J.~Zhou, W.~Fan, and J.~Ye.
\newblock Efficient multi-task feature learning with calibration.
\newblock In \emph{KDD '14}, pages 761--770, 2014.

\bibitem[Hadiji et~al.(2015)Hadiji, Molina, Natarajan, and
  Kersting]{Hadiji2015}
F.~Hadiji, A.~Molina, S.~Natarajan, and K.~Kersting.
\newblock Poisson dependency networks: Gradient boosted models for multivariate
  count data.
\newblock \emph{Machine Learning}, 100\penalty0 (2):\penalty0 477--507, 2015.

\bibitem[Higham(2002)]{Higham:nearPD}
N.~J. Higham.
\newblock Computing the nearest correlation matrix --- a problem from finance.
\newblock \emph{IMA Journal of Numer. Anal.}, 22\penalty0 (3):\penalty0
  329--343, 2002.

\bibitem[Jalali et~al.(2010)Jalali, Sanghavi, Ruan, and
  Ravikumar]{NIPS2010_4125}
A.~Jalali, S.~Sanghavi, C.~Ruan, and P.~K. Ravikumar.
\newblock A dirty model for multi-task learning.
\newblock In \emph{NIPS}, pages 964--972, 2010.

\bibitem[Karlis(2003)]{DKarlis2003}
D.~Karlis.
\newblock An em algorithm for multivariate poisson distribution and related
  models.
\newblock \emph{Journal of Applied Statistics}, 30\penalty0 (1):\penalty0
  63--77, 2003.

\bibitem[Kumar and Daum{\'e}~III(2012)]{conf:icml:KumarD12}
A.~Kumar and H.~Daum{\'e}~III.
\newblock Learning task grouping and overlap in multi-task learning.
\newblock In \emph{ICML '12}, 2012.

\bibitem[Levina et~al.(2008)Levina, Rothman, Zhu, et~al.]{levina2008sparse}
E.~Levina, A.~Rothman, J.~Zhu, et~al.
\newblock Sparse estimation of large covariance matrices via a nested lasso
  penalty.
\newblock \emph{The Annals of Applied Statistics}, 2\penalty0 (1):\penalty0
  245--263, 2008.

\bibitem[Liu et~al.(2014)Liu, Wang, and Zhao]{NIPS2014_5630}
H.~Liu, L.~Wang, and T.~Zhao.
\newblock Multivariate regression with calibration.
\newblock In \emph{NIPS}, pages 127--135, 2014.

\bibitem[Lozano et~al.(2013)Lozano, Jiang, and
  Deng]{Lozano:2013:RSE:2487575.2487667}
A.~C. Lozano, H.~Jiang, and X.~Deng.
\newblock Robust sparse estimation of multiresponse regression and inverse
  covariance matrix via the l2 distance.
\newblock In \emph{KDD '13}, pages 293--301, 2013.

\bibitem[Ma et~al.(2008)Ma, Kockelman, and Damien]{Ma2008:MVPLN_crash}
J.~Ma, K.~M. Kockelman, and P.~Damien.
\newblock {A multivariate Poisson-lognormal regression model for prediction of
  crash counts by severity, using Baysian methods}.
\newblock \emph{Accident Analysis and Prevention}, 40:\penalty0 964--975, 2008.

\bibitem[Rothman et~al.(2010)Rothman, Levina, and Zhu]{Rothman.2010.09188}
A.~J. Rothman, E.~Levina, and J.~Zhu.
\newblock Sparse multivariate regression with covariance estimation.
\newblock \emph{Journal of Computational and Graphical Statistics}, 19\penalty0
  (4):\penalty0 947--962, 2010.

\bibitem[Wang et~al.(2007)Wang, Kalwani, and Akçura]{Deanna:Wang2007369}
H.~Wang, M.~U. Kalwani, and T.~Akçura.
\newblock A bayesian multivariate poisson regression model of cross-category
  store brand purchasing behavior.
\newblock \emph{Journal of Retailing \& Consumer Services}, 14\penalty0
  (6):\penalty0 369--382, 2007.

\bibitem[Wang et~al.(2013)Wang, Liang, and Xing]{conf:aistats:WangLX13}
W.~Wang, Y.~Liang, and E.~P. Xing.
\newblock Block regularized lasso for multivariate multi-response linear
  regression.
\newblock In \emph{AISTATS}, pages 608--617, 2013.

\bibitem[Wang et~al.(2015)Wang, Chakraborty, Mekaru, Brownstein, Ye, and
  Ramakrishnan]{Wang:2015:DPA:2783258.2783291}
Z.~Wang, P.~Chakraborty, S.~R. Mekaru, J.~S. Brownstein, J.~Ye, and
  N.~Ramakrishnan.
\newblock Dynamic poisson autoregression for influenza-like-illness case count
  prediction.
\newblock In \emph{KDD '15}, pages 1285--1294, 2015.

\bibitem[Wytock and Kolter(2013)]{WytockK:icml13}
M.~Wytock and J.~Z. Kolter.
\newblock Sparse gaussian conditional random fields: Algorithms, theory, and
  application to energy forecasting.
\newblock In \emph{ICML '13}, pages 1265--1273, 2013.

\bibitem[Yang et~al.(2013)Yang, Ravikumar, Allen, and Liu]{NIPS2013_5153}
E.~Yang, P.~K. Ravikumar, G.~I. Allen, and Z.~Liu.
\newblock On poisson graphical models.
\newblock In \emph{NIPS '13}. 2013.

\bibitem[Yu et~al.(2007)Yu, Tresp, and Yu]{Yu:2007:RML:1273496.1273635}
S.~Yu, V.~Tresp, and K.~Yu.
\newblock Robust multi-task learning with t-processes.
\newblock In \emph{ICML '07}, 2007.

\end{thebibliography}

\end{document}